\documentclass[aps,prd,10pt,groupedaddress,nofootinbib,floatfix,letterpaper]{revtex4}
\usepackage[utf8]{inputenc}
\usepackage[T1]{fontenc}
\usepackage{lmodern}

\usepackage{dcolumn}
\usepackage{bm}
\usepackage{amssymb}
\usepackage{amsmath}
\usepackage{latexsym}
\usepackage{amsfonts}
\usepackage{tensor}
\usepackage{dsfont}

\usepackage[usenames]{color}
\usepackage{float}
\usepackage{multirow}
\usepackage{epsfig}
\usepackage{graphicx}

\usepackage[caption=false]{subfig} 

\usepackage[colorlinks]{hyperref}

\begin{document}

\title{Nonlinear electrodynamics in Kerr–Newman–NUT–$\Lambda$ spacetime: exact solutions, horizons, and energy conditions}

\author{Oscar Galindo-Uriarte}
\email{oscar.galindo@cinvestav.mx}
\affiliation{Physics Department, Cinvestav, P.O. Box 14-740, Mexico City, Mexico}

\author{Nora Breton}
\email{nora.breton@cinvestav.mx}
\affiliation{Physics Department, Cinvestav, P.O. Box 14-740, Mexico City, Mexico}

\author{Claus Lämmerzahl}
\email{claus.laemmerzahl@zarm.uni-bremen.de}
\affiliation{ZARM, University of Bremen, Am Fallturm, 28359 Bremen, Germany}

\author{Alfredo Macías}
\email{Deceased on November 1st, 2024}
\affiliation{Physics Department, UAM–Iztapalapa, P.O. Box 55–534, C.P. 09340, Mexico City, Mexico}

\begin{abstract}
We construct two exact nonlinear-electrodynamic generalizations of the Kerr–Newman–NUT–$\Lambda$ spacetime. Imposing alignment between the principal directions of the electromagnetic field and the metric tetrad reduces the Maxwell–Faraday sector to a pair of potentials constrained by a single integrability condition, the key equation. Within the polynomial aligned ansatz considered here, the key equation selects two admissible families, corresponding to electromagnetic potentials that are cubic and quartic polynomials. For each family the Einstein equations reduce to a single radial ordinary differential equation that gives a deformation of the Kerr-like radial metric function which is  exactly solved. We derive the corresponding metrics, electromagnetic fields, stress tensors, horizon structure, and we analyze the associated energy conditions. The nonlinear sector breaks conformal invariance and, in the cubic family, can mimic an effective cosmological contribution.  The explicit Lagrangians as functions of the electromagnetic invariants are obtained in selected static subsectors.  The curvature invariants show that the nonlinear contribution does not remove the Kerr-like curvature singularity, while the solutions present the NUT axial conical singularity.

\end{abstract}

\maketitle

\section{Introduction}
It is well established that astrophysical compact objects, such as those at the center of our galaxy, are Kerr-type, that is, very compact rotating objects. In the search for models with these characteristics, incorporating additional physical parameters helps refine their representation, leading to a better understanding and new applications. Topics of great interest include gravitational lensing, black-hole shadows, and gravitational waves. Moreover, the observations of LIGO \cite{LIGO2016} and Virgo \cite{ALAV2021}, as well as star orbits \cite{Ghez2008}, motivate the search for new solutions of Einstein's equations with a black-hole (BH) interpretation.

Independently of BH observations, there are theoretical aspects of black-hole physics that remain unresolved, such as cosmic censorship, the no-hair theorem, the instability of Cauchy horizons, and the nature of BH singularities.

Regarding BH singularities, in the last decade a number of static regular BH solutions have been presented. Some of them are models, i.e. spacetimes that are geometrically regular but whose metrics are not solutions of the Einstein equations \cite{Hayward2006,QtmSch,Lobo2021}; others have been sourced ad hoc with exotic matter. Many of them have been made regular by including nonlinear electromagnetic (NLE) sources, or NLE sources coupled to scalar fields \cite{Saez,Alencar2025}.

Many explicit regular black-hole models are static, while others are obtained through effective rotating prescriptions. Therefore, exact stationary NLED solutions remain of particular interest. In this direction, a method for incorporating NLED sources into stationary Kerr-like geometries was recently developed in Ref.~\cite{GarciaDiaz2022b}.

Although the nonlinear electromagnetic sector considered here does not remove the ring singularity characteristic of the Kerr geometry, the resulting solutions are relevant within the panorama of exact solutions. In particular, exact stationary NLED black holes have only recently been derived in Refs.~\cite{GarciaDiaz2022b,GalindoBreton2024}.

The most relevant stationary axisymmetric electrovacuum solution of the Einstein equations with a BH interpretation is the Kerr-Newman-NUT family \cite{Plebanski1968,Kubiznak}. A nonlinear electromagnetic generalization of the Kerr--Newman--NUT family may therefore shed light on the interplay between the NUT parameter, often interpreted as a dual or magnetic mass, and nonlinear electromagnetic fields. To our knowledge, the only nonlinear electromagnetic generalization in the NUT sector is the static spherically symmetric NUT--Born--Infeld solution, analyzed in Ref.~\cite{Breton2015}, where light trajectories and the extreme case were examined.

In this work, we present two nonlinear electromagnetic black-hole solutions belonging to the Kerr--Newman--NUT--$\Lambda$ family, which we call the cubic and quartic NLE--Kerr--Newman--NUT--$\Lambda$ families, respectively, because the corresponding electromagnetic potentials are cubic and quartic polynomials in coordinates $r$ and $a\cos\theta+n$. These solutions of the Einstein--NLED equations are characterized by seven parameters: mass, angular momentum, electric and magnetic charges, NUT parameter, cosmological constant, and one parameter associated with the nonlinear electromagnetic contribution. The electromagnetic potentials satisfy alignment conditions between the electromagnetic tensor eigenvectors and the metric tetrad.


The procedure has similarities with the standard construction of the Reissner--Nordstr\"om solution: first, one solves the electromagnetic equations, then computes the energy-momentum tensor, and finally solves the corresponding Einstein equations. Here, however, we do not start from a prescribed NLED Lagrangian. Instead, we restrict the set of admissible electromagnetic fields by imposing an alignment condition, motivated by stationarity and axial symmetry, together with an integrability condition. This yields a class of electromagnetic fields that generalize the Kerr--Newman electromagnetic field. With these generalized fields, we compute the energy-momentum tensor and solve the Einstein equations for the corresponding nonlinear metric deformations.

The difference with respect to the usual scheme is that here we are not using a particular field equation derived from a prescribed electromagnetic Lagrangian. Instead, a class of possible electromagnetic fields obeying strong guiding principles, such as the alignment condition and the integrability condition, is constructed. For these fields, the energy-momentum tensor can be calculated and the Einstein equations can be solved. The NLED Lagrangian can be calculated in principle \cite{GarciaDiaz2022b}, but a closed global expression $\mathcal{L}(F,G)$ is not obtained for the rotating families. Instead, we reconstruct the on-shell Lagrangian density explicitly and exhibit invariant forms in selected static subsectors.

The paper is organized as follows. Section~\ref{NLED-gravity Field equations} presents the Einstein--NLED field equations. Section~\ref{The Kerr-Newman-NUT-Lambda metric} reviews the Kerr--Newman--NUT--$\Lambda$ spacetime. Section~\ref{Method for generating NLED rotating solutions} describes the aligned reconstruction method and derives the key equation and the radial master equation. Sections~\ref{CubicSol} and~\ref{QuarticSol} present the cubic and quartic NLED families of the Kerr--Newman--NUT--$\Lambda$ metric, respectively, their horizon structure and degenerate limits. Section~\ref{EnergyT} analyzes the canonical stress tensor, the energy conditions, the breaking of conformal invariance, and the corresponding on-shell Lagrangians. Section~\ref{Conclusions} contains  final remarks. Appendices~\ref{KretS} and~\ref{kret.quartic} collect the full Kretschmann scalars, Appendix~\ref{app:Lagrangian} discusses the on-shell Lagrangians and static invariant limits, and Appendix~\ref{app:roots} summarizes the quartic root formulas used in the horizon analysis.


\section{NLED-gravity Field equations}\label{NLED-gravity Field equations}

The action for nonlinear electrodynamics (NLED) minimally coupled to Einstein gravity is given by \cite{Salazar1987}
\begin{equation}
W = \int d^4x \, \sqrt{-g} \left\{ \frac{R - 2 \Lambda}{16\pi} + \mathcal{L}_{\text{E}}(F,G) \right\},
\end{equation}
where $g$ is the determinant of the metric tensor $g_{\mu \nu}$ with signature $(- + + +)$, $R$ is the Ricci scalar, $\Lambda$ is the cosmological constant, and $\mathcal{L}_{\text{E}}(F,G)$ denotes the nonlinear electromagnetic Lagrangian density. The electromagnetic invariants $F$ and $G$ are defined as
\begin{align}
F &:= \tfrac{1}{4} F_{\mu\nu} F^{\mu\nu},&
G &:= \tfrac{1}{4} F_{\mu\nu} \,{}^{\star}\!F^{\mu\nu},
\label{invFG}
\end{align}
where the dual electromagnetic tensor is defined \citep[Eq.~(3.51)]{MTW1973} as
\begin{align}\label{duals}
{}^\star{F}_{\mu \nu} &= \tfrac{1}{2}\sqrt{-g}\,\varepsilon_{\mu \nu \alpha\beta}F^{\alpha \beta},
&\quad {}^{\star}F^{\mu\nu} &= -\tfrac{1}{2}\frac{\varepsilon^{\mu \nu \alpha\beta}}{\sqrt{-g}}F_{\alpha \beta},
\end{align}
where $\varepsilon^{\mu \nu \alpha\beta}$ is the Levi-Civita symbol and $g = \det(g_{\mu\nu})$.
Note that the double dual satisfies ${}^\star\!\left({}^\star{F}_{\mu \nu}\right) = -F_{\mu \nu}$.
In terms of the electric field and magnetic induction, $\vec E$ and $\vec B$, the invariants $F$ and $G$ are given by
\begin{align}
F &:= \tfrac{1}{2}\!\left(\vec B^{\,2}-\vec E^{\,2}\right), &
G &:= -\,\vec E\cdot\vec B, \label{invFG2}
\end{align}
where
\begin{align}
E^i &= F^{0i}, &
B^i &= \tfrac{1}{2}\,\epsilon^{ijk} F_{jk},
\end{align}
so that $\vec E^{\,2}:=\delta_{ij}E^iE^j$, and analogously for $\vec B$; $i, j = 1,2,3$.
The electromagnetic tensor is given in terms of the electromagnetic potential $A_{\mu}$ as $F_{\mu\nu}=\partial_\mu A_\nu-\partial_\nu A_\mu$.

It will be convenient to introduce the rescaled Lagrangian
\begin{equation}
\mathcal L(F,G):=-4\pi\,\mathcal L_{\rm E}(F,G).
\label{def:rescaled-structural-function}
\end{equation}
All derivatives $\mathcal L_F$ and $\mathcal L_G$ used below refer to this rescaled function,
\begin{equation}
\mathcal L_F:=\frac{\partial \mathcal L}{\partial F},
\qquad
\mathcal L_G:=\frac{\partial \mathcal L}{\partial G}.
\end{equation}
With this convention, the Maxwell limit is
\begin{equation}
\mathcal L_{\rm Maxwell}=-F,
\qquad
\mathcal L_{{\rm E},\,{\rm Maxwell}}=\frac{F}{4\pi}.
\label{Maxwell-limit-convention}
\end{equation}

Variation of the action with respect to the electromagnetic potential $A_{\mu}$ yields the generalized Maxwell equations in absence of sources,
\begin{equation}\label{FM}
\nabla^{\nu} \left( \mathcal{L}_F F_{\mu\nu} + \mathcal{L}_{G} \, {}^{\star}F_{\mu\nu} \right) = 0,
\end{equation}
where  $\nabla_\nu$ means the covariant derivative.
It is convenient to introduce the antisymmetric Plebanski tensor
\begin{equation}
P_{\mu\nu} = \mathcal{L}_F F_{\mu\nu} + \mathcal{L}_{G} \, {}^{\star}F_{\mu\nu},
\label{constEq}
\end{equation}
so that Eq.~\eqref{FM} formally resembles Maxwell’s equations under the replacement $F_{\mu\nu} \mapsto P_{\mu\nu}$.

The generalized Maxwell equations must be supplemented by the Bianchi identities. Locally, these imply that $F$ and ${}^\star P$ can be written in terms of potentials:
\begin{align}
\label{Maxwell.Eq} \nabla_\nu P^{\mu\nu} &= 0
\quad \longleftrightarrow \quad {}^{\star}P_{[\mu\nu;\gamma]}=0
\quad \longleftrightarrow \quad {}^\star{P}_{\mu\nu}=\partial_\mu \tilde{\mathcal{P}}_\nu-\partial_\nu \tilde{\mathcal{P}}_\mu, \\
\label{Faraday.Eq}\nabla_\nu {}^{\star}F^{\mu\nu} &= 0
\quad \longleftrightarrow \quad F_{[\mu\nu;\gamma]}=0
\quad \longleftrightarrow \quad F_{\mu\nu}=\partial_\mu A_\nu-\partial_\nu A_\mu.
\end{align}

In the linear Maxwell limit, $P_{\mu\nu}$ becomes proportional to $F_{\mu\nu}$, and the standard source-free Maxwell equations are recovered. Variation of the action with respect to the metric $g^{\mu \nu}$ leads to the Einstein equations,
\begin{equation}\label{eq.Einsten}
G_{\mu\nu} + \Lambda g_{\mu\nu} = R_{\mu\nu} - \tfrac{1}{2} R g_{\mu\nu} + \Lambda g_{\mu\nu} = 8\pi T_{\mu\nu},
\end{equation}
where $G_{\mu\nu} $ is the Einstein tensor, $R_{\mu\nu} $ is the Ricci tensor, $R$ is the Ricci scalar and $T_{\mu\nu}$ is the energy–momentum tensor.
In here we shall use the Plebanski tensor definition \citep[Eq.~(2.3b)]{Salazar1987}
\begin{equation}\label{tmunu.NLE}
 \begin{split}
     T_{\mu\nu}&=-\frac{2}{\sqrt{-g}}\frac{\delta(\sqrt{-g}\mathcal{L}_{\text{E}}(F,G))}{\delta g^{\mu\nu}}= g_{\mu\nu} \mathcal{L}_{\text{E}}(F,G)-2\frac{\delta \mathcal{L}_{\text{E}}(F,G)}{\delta g^{\mu\nu}},\\
4\pi T_{\mu\nu}&= g_{\mu\nu}\,\mathcal{L} -\tensor{P}{_\mu^\alpha}\tensor{F}{_\nu_\alpha},
\qquad \mathcal{L}_{\text{E}}(F,G) = -\frac{\mathcal{L}}{4\pi}.
 \end{split}
\end{equation}

Contracting indices gives
\begin{equation}
\pi {T^\mu}_\mu
=
\mathcal L
-
\frac{1}{4}P^{\mu\nu}F_{\mu\nu},
\label{trace:NLED}
\end{equation}
or, equivalently,
\begin{equation}
\mathcal L
=
\frac{1}{4}P^{\mu\nu}F_{\mu\nu}
+
\pi {T^\mu}_\mu.
\label{L-from-trace}
\end{equation}


\section{\texorpdfstring{The Kerr-Newman-NUT-$\Lambda$ metric}{The Kerr-Newman-NUT-Lambda metric}}\label{The Kerr-Newman-NUT-Lambda metric}

For completeness in this Section we review the linear electromagnetic case, that is, the stationary axisymmetric solution of the coupled Einstein-Maxwell equations, 
the Kerr-Newman-NUT-$\Lambda$ (KN-NUT-$\Lambda$) metric, which in Boyer-Lindquist coordinates \cite{Griffiths2006} is given by
\begin{equation}\label{metric1}
\begin{split}
ds^2&=-{\frac{\Delta_{r}}{\Sigma \Xi^2}}\left( {{dt}} -
\chi{ d \phi} \right)^2 + \frac{\Sigma}{\Delta_{r}}\,{ dr}^2 +\frac{\Sigma}{\Delta_{\theta}}\,{ d \theta}^2+ \frac{\sin^2{\theta} \Delta_{\theta}}{\Sigma \Xi^2} \left[ {a {dt}}
-\left(\Sigma+a\chi\right) { d \phi} \right]^2,
\end{split}
\end{equation}
the signature used is $(- + + +)$ and the metric functions are defined as
\begin{equation}\label{Delta.KN}
\begin{split}
\Delta^{KN}_{r}(r) &:=r^2-2mr-\frac{\Lambda r^2}{3}(r^2+a^2+6n^2)+(1-n^2\Lambda)(a^2-n^2)+Q_e^2+Q_m^2,\\
 \Delta_{\theta} (\theta) &:= 1 + \frac{\Lambda}{3}a \cos \theta\left(a\cos\theta+4n\right),\\
 \Sigma(\theta,r)&:={{r^2+(a\cos\theta+n)^2}};
 \quad\chi:=a \sin^2 \theta+2n(1-\cos\theta);\quad\Xi=1+\frac{\Lambda}{3}a^2,\\
 \Sigma+a\chi&=r^2+(a+n)^2.
\end{split}
\end{equation}
The parameters are angular momentum $a$, mass $m$, cosmological constant $\Lambda$ that may be positive  (de Sitter) or negative (anti de Sitter), NUT parameter $n$,and $Q_e$ and $Q_m$ being the BH electric and magnetic charges, respectively.

\subsection{\texorpdfstring{The KN-NUT-$\Lambda$ electromagnetic field.}{The KN-NUT-Lambda electromagnetic field.}}

The Lagrangian in the linear case is $\mathcal{L}^{KN}=-F$; and the KN-NUT-$\Lambda$  electromagnetic potentials are given by
\begin{equation}\label{Amu.KN.NUT}
\begin{split}
    A^{KN}_\mu dx^\mu&={-\frac{Q_e r}{\Sigma \Xi}}\left( {{dt}} -\chi{ d \phi} \right)+\frac{Q_m (a\cos {\theta}+n)}{a\Sigma \Xi} \left[ {a {dt}}-\left(\Sigma+a\chi\right) { d \phi} \right]
\end{split}
\end{equation}
and
\begin{equation}\label{pmu.KN.NUT}
\begin{split}
   \tilde{\mathcal{P}}^{KN}_\mu dx^\mu&={\frac{Q_m r}{\Sigma \Xi}}\left( {{dt}} -\chi{ d \phi} \right)+\frac{Q_e (a\cos {\theta}+n)}{a\Sigma \Xi} \left[ {a {dt}}-\left(\Sigma+a\chi\right) {d \phi} \right].
\end{split}
\end{equation}

From (\ref{Amu.KN.NUT}) and (\ref{pmu.KN.NUT}), we obtain the electromagnetic field tensor
\begin{align}\label{Fmn.KN.NUT}
F_{\mu\nu} dx^\mu \wedge dx^\nu&=F_{r t}^{KN} {{dr}}\wedge\left( {{dt}} -\chi{ d \phi} \right)+\frac{F_{\theta t}^{KN} }{a}{{d\theta}} \wedge\left[ {a {dt}}-\left(\Sigma+a\chi\right) { d \phi} \right],\\\label{Pmn.KN.NUT}
{}^{\star}P_{\mu\nu} dx^\mu \wedge dx^\nu&={}^{\star}P_{r t}^{KN} {{dr}}\wedge\left( {{dt}} -\chi{ d \phi} \right)+\frac{{}^{\star}P_{\theta t}^{KN} }{a}{{d\theta}} \wedge\left[ {a {dt}}-\left(\Sigma+a\chi\right) { d \phi} \right].
\end{align}

Note that the nonvanishing components of $F_{\mu \nu}^{KN}$ are $F_{r t}^{KN}$, $F_{r \phi}^{KN}$, $F_{\theta t}^{KN}$ and $F_{\theta \phi}^{KN}$ and they are related by
\begin{equation}
F_{r \phi}^{KN}= - \chi F_{r t}^{KN}, \quad a F_{\theta \phi }^{KN}= - (\Sigma+a\chi)F_{\theta t}^{KN},    
\end{equation}
such that there are only two independent components. The same  applies to the components of the tensor
${}^{\star}P_{\mu \nu}^{KN}$.
Explicitly,  the Kerr-Newman-NUT (KN) components are given by,
\begin{align}\label{Fmunu.KN}
F_{r t}^{KN} &=  - \frac{2 Q_m r }{\Xi\Sigma^2}(a\cos {\theta}+n)+\frac{Q_e}{\Xi\Sigma^2}\left[r^2-(a\cos\theta+n)^2\right],\\
F_{\theta t}^{KN} &=-  \frac{2 Q_e\,r a \sin {\theta}}{\Xi\Sigma^2}(a\cos {\theta}+n)-\frac{a Q_m\sin {\theta}}{\Xi\Sigma^2}  \left[r^2- (a\cos\theta+n)^2\right],\\
{}^{\star}P_{r t}^{KN} &= - \frac{2 Q_e r }{\Xi\Sigma^2}(a\cos {\theta}+n)-\frac{Q_m}{\Xi\Sigma^2}\left[r^2-(a\cos\theta+n)^2\right],\\
{}^{\star}P_{\theta t}^{KN} &=  \frac{2 Q_m\,r a \sin {\theta}}{\Xi\Sigma^2}(a\cos {\theta}+n)-\frac{a Q_e\sin {\theta}}{\Xi\Sigma^2}  \left[r^2- (a\cos\theta+n)^2\right].
\end{align}

\subsection{\texorpdfstring{The KN-NUT-$\Lambda$ horizons.}{The KN-NUT-Lambda horizons.}}
In stationary, axisymmetric solutions such as the Kerr–Newman–NUT-$\Lambda$ spacetime, the definition of the event horizon corresponds to a null hypersurface located at $r=r_{+}$, where the normal vector becomes null. In Boyer--Lindquist-like coordinates the Kerr-like metric component $g^{rr} $ is
\begin{equation}
g^{rr} = \frac{\Delta_r}{\Sigma},
\end{equation}
such that the condition $\Delta_r(r_{+})=0$ determines the horizon.
Geometrically, this means that across $r=r_{+}$ the hypersurface changes causal character, with $g^{rr}$ switching sign from positive to negative, thereby signaling the presence of the event horizon.


In case $\Lambda=0$ there are two horizons given by the solutions of $\Delta^{KN}_{r}=0$,

\begin{equation}
r^{KN}_{\pm} = m \pm \sqrt{m^2- \left( Q^2+a^2-n^2 \right)},
\end{equation}
where $Q^2=Q_e^2+Q_m^2$. Note that there is a bound for the parameters
$a$, $Q$ and $n$, in order that the square root be positive,
\begin{equation}
m^2 \ge Q^2 + a^2 - n^2.
\end{equation}

The Kerr-Newman-NUT solution of the Einstein-Maxwell equations belongs to the Pleba\'nski class, which admits a group $G_2$ with two commuting Killing vectors and was studied in detail in \cite{Plebanski1975}. The Pleba\'nski class includes several interesting cases, as shown in Table \ref{tab:spacetimes}; the NLED generalizations presented in this work belong as well to the Pleba\'nski class, since they share the same metrical structure
In case $a=0$ and $\Lambda=0$ the Reissner-Nordstrom-NUT metric is obtained; this is the static case of the Einstein-Maxwell equations, equipped with mass $m$, NUT parameter $n$, and electric and/or magnetic charges,
 $Q_e$, $Q_m$.
The area  of the event horizon hypersurface depends also on the NUT parameter; for a surface with $t=$const. and $r=r_{+}$, where $r_{+}$ is the horizon radius, reads
\begin{equation}
A = \int_{0}^{2 \pi} d \phi \int_{0}^{ \pi} d \theta \sqrt{|g_{\theta \theta} g_{\phi \phi}|} = 4 \pi \frac{[r_{+}^2 +(a+n)^2]}{\Xi}.
\end{equation}
It is well known that $A$ is related to the entropy of the BH, which is larger if $n \ne 0$; then the thermodynamical quantities are also modified by the presence of the NUT parameter.

This geometry can be extended to include NLED matter, comprised in additional NLED parameters,  $\beta$ and $\xi$, indicating two types of non-linear extensions. In the next section we review the method to include NLED into the Kerr-like geometry.

\section{Method for generating NLED rotating solutions}\label{Method for generating NLED rotating solutions}

In this section we outline the method used to generate the NLED generalizations of Kerr-like rotating geometries. The procedure relies on the so-called \textit{alignment conditions}, which correspond to choosing an appropriate tetrad basis such that the mixed electromagnetic tensor $\tensor{F}{^{a}_{b}}$ becomes diagonal in that frame. Since $F_{ab}$ is antisymmetric, it cannot be diagonalized as a two-form. However, when the tetrad index is raised, the mixed tensor $\tensor{F}{^{a}_{b}}$ has its eigenvalues on the diagonal provided that the tetrad vectors are aligned with the eigenvectors of the electromagnetic field. In this sense, the alignment conditions enforce a geometric compatibility between the electromagnetic field and the underlying Kerr-like symmetries.
Once the general form of the electromagnetic potentials $A_{\mu}$ and $\tilde{\mathcal{P}}_{\mu}$ is fixed under these conditions, the potentials must satisfy the Lagrangian integrability constraints, which ultimately lead to the so-called \textit{key equation}. This equation restricts arbitrary functions in the electromagnetic potentials. The complete procedure is summarized below. 
On the other hand,
the metric function $\Delta^{KN}_{r}$ is generalized through
${\Delta}_{r} \mapsto \Delta^{KN}_{r} + f(r)$, where $f(r)$ modifies the metric to include the nonlinear electromagnetic contribution, and it is determined from the Einstein-NLED equations. 
\subsection{Alignment conditions}

The alignment conditions were originally introduced in the context of stationary and axisymmetric nonlinear electrodynamics black hole solutions \cite{Garcia1984_BornInfeld}, with further developments in \cite{GarciaDiaz2022b,GalindoBreton2024,AyonBeato2024}. These conditions arise from requiring that the complex two-form
\begin{equation}
\omega = \tfrac{1}{2} \left( F_{\mu\nu} + i {}^{\star} P_{\mu\nu} \right) dx^{\mu} \wedge dx^{\nu},
\label{Kerr_omega}
\end{equation}
possess only four nonvanishing components compatible with stationarity and axial symmetry: $F_{\theta\phi},~F_{\theta t},~F_{r\phi},~F_{rt}$. The closure condition of the two-form $\omega$ is equivalent to the Maxwell and Faraday equation

\begin{equation}
    d\omega=0,\leftrightarrow \nabla_\nu P^{\mu\nu} = 0, \quad \nabla_\nu {}^{\star}F^{\mu\nu} = 0,
    \end{equation}
respectively. In the Newman-Penrose null tetrad formalism \cite{Plebanski1975}, the Kerr-like geometry can be expressed as
\begin{equation}
ds^{2}=2\,\mathbf{e^1 \oplus e^2} + 2\,\mathbf{e^3 \oplus e^4},\quad \mathbf{e^1}=\overline{\mathbf{e^2}},\quad \mathbf{e^3}=\overline{\mathbf{e^3}},\quad \mathbf{e^4}=\overline{\mathbf{e^4}},
\end{equation}
where the null tetrad is given by
\begin{align}
e^{1,2} &= \frac{1}{\sqrt{2}}
\left\{
\sqrt{\frac{\Sigma}{\Delta_\theta}}\, \mathbf{d\theta}
\pm i \frac{\sin\theta}{\Xi}\sqrt{\frac{\Delta_\theta}{\Sigma}}
\left[ a\,\mathbf{dt} - (\Sigma + a\chi)\,\mathbf{d\phi} \right]
\right\},\\[2mm]
e^{3,4} &= \frac{1}{\sqrt{2}}
\left\{
\sqrt{\frac{\Sigma}{\Delta_r}}\, \mathbf{dr}
\pm 
\sqrt{\frac{\Delta_r}{\Sigma}}
\left( \frac{\mathbf{dt} - \chi\,\mathbf{d\phi}}{\Xi} \right)
\right\}.
\end{align}

In this tetrad, that we denote with a capital $N$ for null, the two-form $NF_{ab}$ takes the canonical antisymmetric form
\begin{equation}
NF_{ab} =
\begin{pmatrix}
0 & \displaystyle \frac{i\Xi}{a \sin\theta}\frac{\partial A_t}{\partial \theta} & 0 & 0 \\[3mm]
-\displaystyle \frac{i\Xi}{a \sin\theta}\frac{\partial A_t}{\partial \theta} & 0 & 0 & 0 \\[3mm]
0 & 0 & 0 &\displaystyle \Xi \frac{\partial A_t}{\partial r} \\[3mm]
0 &0 &-\displaystyle \Xi \frac{\partial A_t}{\partial r} & 0
\end{pmatrix}.
\end{equation}

Raising one tetrad index yields the mixed tensor
\begin{equation}
\tensor{NF}{^\mu_\nu} =
\begin{pmatrix}
-\displaystyle \frac{i\Xi}{a\sin\theta}\frac{\partial A_t}{\partial \theta} & 0 &0 & 0\\[3mm]
0 & \displaystyle \frac{i\Xi}{a\sin\theta}\frac{\partial A_t}{\partial \theta} & 0 &0 \\[3mm]
0 & 0 & \displaystyle \Xi\frac{\partial A_t}{\partial r} & 0\\[3mm]
0 &0 & 0 & -\displaystyle \Xi\frac{\partial A_t}{\partial r}
\end{pmatrix},
\end{equation}
which is diagonal and whose entries correspond to the eigenvalues of $F_{\mu\nu}$.
This is precisely what is meant by alignment: the tetrad is chosen so that its vectors coincide with the eigenvectors of the electromagnetic field.

Under these conditions, only two independent components of the electromagnetic potentials exist: $A_t$ and $\tilde{\mathcal{P}}_t$. The alignment conditions impose the relations
\begin{align}\label{AC.At.r}
\frac{\partial A_\phi}{\partial r} &= -\chi \frac{\partial A_t}{\partial r}, \nonumber\\
\frac{\partial \tilde{\mathcal{P}}_\phi}{\partial r} &= -\chi \frac{\partial \tilde{\mathcal{P}}_t}{\partial r}, \nonumber\\
a\frac{\partial A_\phi}{\partial \theta} &= -(\Sigma + a\chi)\frac{\partial A_t}{\partial \theta}, \nonumber\\
a\frac{\partial \tilde{\mathcal{P}}_\phi}{\partial \theta} &= -(\Sigma + a\chi)\frac{\partial \tilde{\mathcal{P}}_t}{\partial \theta}.
\end{align}

Equivalently, for the field components:
\begin{align}\label{AC.At.rfields}
F_{r\phi} &= -\chi\, F_{rt}, \nonumber\\
{}^{\star}P_{r\phi} &= -\chi\, {}^{\star}P_{rt}, \nonumber\\
aF_{\theta\phi} &= -(\Sigma + a\chi)F_{\theta t}, \nonumber\\
a\, {}^{\star}P_{\theta\phi} &= -(\Sigma + a\chi)\,{}^{\star}P_{\theta t}.
\end{align}

The general solution of Eqs.~(\ref{AC.At.r}) is
\begin{align}
A_t &=\frac{X(r)+Y(\theta)}{\Sigma}, \qquad
A_\phi = -\chi\,\frac{X(r)}{\Sigma} - \frac{\Sigma + a\chi}{a} \frac{Y(\theta)}{\Sigma},
\label{GS.At}\\[2mm]
\tilde{\mathcal{P}}_t &=\frac{A(r)+B(\theta)}{\Sigma}, \qquad
\tilde{\mathcal{P}}_\phi = -\chi\,\frac{A(r)}{\Sigma} - \frac{\Sigma + a\chi}{a} \frac{B(\theta)}{\Sigma},
\label{GS.Pt}
\end{align}
where $X(r)$, $Y(\theta)$, $A(r)$, and $B(\theta)$ are  functions 
to be determined from the Lagrangian integrability constraints, which ultimately lead to the so-called \textit{key equation} which restricts the initially arbitrary functions $X(r)$, $Y(\theta)$, $A(r)$, and $B(\theta)$. The complete procedure is described in the following. 
 On the other hand,
the metric function ${\Delta}_{r}^{KN}$ is generalized as follows
\begin{equation}
{\Delta}_{r} ={\Delta}_{r}^{KN} + f(r) = r^2-2mr-\frac{\Lambda r^2}{3}(r^2+a^2+6n^2)+(1-n^2\Lambda)(a^2-n^2)+Q_e^2+Q_m^2 + f(r),    
\end{equation}
where $f(r)$ modifies the KN metric to include the nonlinear electromagnetic contribution that is determined from the Einstein-NLED equations.
\subsection{Nonlinear electromagnetic Lagrangian and the trace of the energy-momentum tensor}

We shall now consider a
Kerr-Newman-NUT-$\Lambda$-like metric in Boyer-Lindquist coordinates
\cite{Griffiths2006} is

\begin{equation}\label{metric2}
\begin{split}
ds^2&=-{\frac{{\Delta}_{r}}{\Sigma \Xi^2}}\left( {{dt}} -
\chi{ d \phi} \right)^2 + \frac{\Sigma}{{\Delta}_{r}}\,{ dr}^2 +\frac{\Sigma}{\Delta_{\theta}}\,{ d \theta}^2+ \frac{\sin^2{\theta} \Delta_{\theta}}{\Sigma \Xi^2}  \left[ {a {dt}}
-\left(\Sigma+a\chi\right) { d \phi} \right]^2,
\end{split}
\end{equation}
where ${\Delta}_{r} = \Delta^{KN}_{r}+ f(r)$,  $\Delta^{KN}_{r}$ is the Kerr-Newman-NUT-$\Lambda$ metric function in Eq. (\ref{metric1}) and  $f(r)$ is to be determined from Einstein equations (\ref{eq.Einsten}).
From the left hand side of
$G_{\mu\nu}+\Lambda g_{\mu\nu}=8\pi T_{\mu\nu}$, the trace of the energy-momentum tensor can be determined as

\begin{align}
G^{\mu}_{\mu} + 4 \Lambda &=  \frac{1}{\Sigma}\left(\frac{\partial^2 \Delta_r}{\partial r^2}-2\Delta_\theta+3\frac{\cos\theta}{\sin\theta}\frac{\partial \Delta_\theta}{\partial \theta}+\frac{\partial^2\Delta_\theta}{\partial \theta^2}\right)+4\Lambda= \frac{1}{\Sigma} \left[ \Delta_{r}{}^{\prime \prime} - 2 + \left(\frac{2}{3}a^2+4r^2+4n^2 \right)\Lambda \right] = 8 \pi \tensor{T}{^\mu_\mu}.
\end{align}

While from the electromagnetic stress-energy tensor, $4\pi T_{\mu\nu}=\mathcal{L} g_{\mu\nu}-{F}_{\mu .}^{\alpha} P_{\nu\alpha}$, the expression for the Lagrangian $\mathcal{L}$ is,
\begin{equation}
    \mathcal{L}=\frac{1}{4}\ {F}^{\mu\nu} P_{\mu \nu} +\pi {T}{^\mu_\mu} 
= -\frac{\Xi^2}{2a\sin\theta} \left(\frac{\partial A_t}{\partial r}\frac{\partial \tilde{\mathcal{P}}_t}{\partial \theta} + \frac{\partial A_t}{\partial \theta} \frac{\partial \tilde{\mathcal{P}}_t}{\partial r}\right)+\frac{1}{8\Sigma} \left[ \Delta_{r}{}^{\prime \prime} - 2 + \left( \frac{2}{3}a^2+4 r^2+4 n^2 \right)\Lambda \right].
\label{LagrPots}
\end{equation}

Identifying the electromagnetic intensities electric field $E$, magnetic induction $B$, electric displacement $D$, and magnetic field $H$—as in \citep[eq.~(20b)]{AyonBeato2024},
\begin{equation}\label{Field.def}
    E=-\Xi\frac{\partial A_t}{\partial r},\quad
    B=-\frac{\Xi}{a\sin\theta}\frac{\partial A_t}{\partial \theta},\quad
    D=\frac{\Xi}{a\sin\theta}\frac{\partial \tilde{\mathcal{P}}_t}{\partial \theta}, \quad
    H=-\Xi\frac{\partial \tilde{\mathcal{P}}_t}{\partial r}.
\end{equation}
The potentials are written in a gauge adapted to the rotating case $(a\neq0)$. Static limits are understood either at the level of the field strengths or after an appropriate gauge redefinition, so that the apparent ($1/a$) factors do not represent physical singularities. The corresponding field components are written as
\begin{equation}\label{FP.components}
F_{rt}=-\frac{E}{\Xi},\quad
F_{\theta t}=-\frac{a\sin\theta}{\Xi}B,\quad
{}^{\star}P_{rt}=-\frac{H}{\Xi},\quad
{}^{\star}P_{\theta t}=\frac{a\sin\theta}{\Xi}D.
\end{equation}

Thus, the matter field Lagrangian reduces to
\begin{equation}\label{lag.matter}
    \mathcal{L}=\frac{ED-BH}{2}+\frac{1}{8\Sigma} \left[ \Delta_{r}{}^{\prime \prime} - 2 + \left(\tfrac{2}{3}a^2+4r^2+4n^2 \right)\Lambda \right].
\end{equation}
\subsection{\texorpdfstring{The linear electromagnetic case, Kerr-Newman-NUT-$\Lambda$}{} }
For the sake of illustration we review the linear case of Section \ref{The Kerr-Newman-NUT-Lambda metric} for the Lagrangian $\mathcal{L}^{KN}=-F$,
corresponding to the Kerr-Newman-NUT-$\Lambda$ solution;
$\mathcal{L}^{KN}$ is  given by
\begin{equation}
\mathcal{L}^{KN}= -F=\frac{1}{2} (E^2-B^2)  = \frac{1}{2} \left[  (F_{r t}^{KN})^2-\left(\frac{F_{\theta  t}^{KN}}{a \sin \theta} \right)^2 \right],    
\end{equation}
and the components  $F_{\theta t}^{KN}$ and  $F_{r t}^{KN}$ are given in Eqs. (\ref{Fmunu.KN}). For the Kerr-Newman-NUT-$\Lambda$ solution   the second term in the Lagrangian Eq. (\ref{LagrPots}) or (\ref{lag.matter}) vanishes as it corresponds to the null trace of $T_{\mu \nu}$ of KN-NUT-$\Lambda$ solution, with $\Delta_{r} = \Delta^{KN}_{r}$ given in Eqs. (\ref{metric1}). 
Particularly, the Lagrangian for the Kerr-Newman-NUT-$\Lambda$ metric in terms of coordinates $(r, \theta)$ is given by
\begin{equation}
\mathcal{L}_{\text{KN-NUT}} = \frac{h_1(r,\theta) h_2(r,\theta)}{2\Sigma^4}\, ,
\label{KN_Lagr}
\end{equation}
where the functions $h_1(r,\theta)$ and $h_2(r,\theta)$ are given by
\begin{equation}
\begin{split}
h_1(r,\theta)&=(Q_e+Q_m)\left[r^2-(a\cos\theta+n)^2\right]+2r(Q_e-Q_m)(a\cos\theta+n), \\
h_2(r,\theta)&=(Q_e-Q_m)\left[r^2-(a\cos\theta+n)^2\right]-2r(Q_e+Q_m)(a\cos\theta+n).
\end{split}
\label{h1h2}
\end{equation}
It is illustrative to present the expression of the Lagrangian of the Kerr-Newman-NUT solution with an electric charge only, $Q_m=0$,
\begin{equation}
\mathcal{L}_{\text{KN-NUT}} =  \frac{Q_e^2}{2 \Sigma^4} \left[ r^4-6 r^2 (n + a \cos \theta)^2 + (n + a \cos \theta)^4 \right]
\end{equation}


\begin{table}[H]
\centering
\begin{tabular}{|l|c|}
\hline
Spacetime & Parameters \\ \hline
Schwarzschild & $m$ \\
Reissner--Nordström (RN) & $m, Q$ \\
Kerr & $m, a$ \\
Kerr--Newman (KN) & $m, a, Q$ \\
Schwarzschild-$\Lambda$ & $m, \Lambda$ \\
RN-$\Lambda$ & $m, \Lambda, Q$ \\
Kerr-$\Lambda$ & $m, \Lambda, a$ \\
KN-$\Lambda$ & $m, \Lambda, a, Q$ \\
NUT & $m, n$ \\
RN-NUT & $m, n, Q$ \\
Kerr-NUT & $m, n, a$ \\
KN-NUT & $m, n, a, Q$ \\
$\Lambda$-NUT & $m, n, \Lambda$ \\
RN-$\Lambda$-NUT & $m, n, \Lambda, Q$ \\
Kerr-$\Lambda$-NUT & $m, n, \Lambda, a$ \\
KN-$\Lambda$-NUT & $m, n, \Lambda, a, Q$ \\
RN-NLE & $m, Q, \beta$ \\
KN-NLE & $m, a, Q, \beta$ \\
RN-$\Lambda$-NLE & $m, \Lambda, Q, \beta$ \\
KN-$\Lambda$-NLE & $m, \Lambda, a, Q, \beta$ \\
RN-NUT-NLE & $m, n, Q, \beta$ \\
KN-NUT-NLE & $m, n, a, Q, \beta$ \\
RN-$\Lambda$-NUT-NLE & $m, n, \Lambda, Q, \beta$ \\
KN-$\Lambda$-NUT-NLE & $m, n, \Lambda, a, Q, \beta$ \\ \hline
\end{tabular}
\caption{Classification of the 24 special cases encompassed in the nonlinear electrodynamics (NLED) generalization of the KN–NUT–$\Lambda$ spacetime.
The parameters are: mass $m$; NUT parameter $n$; cosmological constant $\Lambda$; rotation parameter $a$; electric/magnetic charge $Q$; and the NLE coupling $\beta$ for the Cubic NLED family and $\xi$ for the Quartic NLED one.}
\label{tab:spacetimes}
\end{table}
\subsection{\texorpdfstring{The key equation for the NLE potentials $A_{\mu}$ and ${}^{\star} P_{\mu}$}{The key equation for the NLE potentials A\_mu and *P\_mu}}
The electromagnetic potentials $A_{\mu}$ and $\tilde{\mathcal{P}}_{\mu}$ in Eqs. (\ref{GS.At})-(\ref{GS.Pt}) are not completely arbitrary, but rather they are constrained by the constitutive Eq. (\ref{constEq}) that links $P_{\mu\nu}$ and $F_{\mu\nu}$. In addition, they must satisfy the integrability (or closure) condition of the Lagrangian,
\begin{equation}\label{closureL}
\frac{\partial^2  \mathcal{L}}{\partial {\theta}  \partial r}= \frac{\partial^2  \mathcal{L}}{\partial {r}  \partial {\theta}},
\end{equation}
which guarantees the compatibility of the field equations.

Following the aligned reconstruction approach, introduced in \cite{GarciaDiaz2022b} and further developed in \cite{GalindoBreton2024}, 
the Lagrangian is taken as a function of $(E,B)$ (provided that the mapping $(r,\theta)\mapsto(E,B)$ is invertible in a given branch, this representation can be translated into invariant variables), or equivalently of the invariants $(F,G)$~\citep[eq.~(2.23)]{Salazar1987}, so that
\begin{equation}
     d\mathcal{L}=-D\, dE+H\, dB,
\end{equation}
together with the relation (\ref{Field.def}). 
The closure condition (\ref{closureL}) then takes the form
\begin{equation}\label{key.eq.fields}
\frac{\partial E}{\partial r}\frac{\partial D}{\partial \theta}+\frac{\partial B}{\partial \theta}\frac{\partial H}{\partial r} =0.
\end{equation}

Explicitly, this can be written as
\begin{equation}\label{key.eq}
\left( \frac{\partial^2 A_{t}}{\partial r^2} \right)
\frac{\partial}{\partial {\theta}} \left( \frac{1}{\sin{\theta}} \frac{\partial \tilde{\mathcal{P}}_{t}}{\partial {\theta}}\right)
-\left( \frac{\partial^2 \tilde{\mathcal{P}}_{t}}{\partial r^2} \right)
\frac{\partial}{\partial {\theta}} \left( \frac{1}{\sin{\theta}} \frac{\partial A_{t}}{\partial {\theta}}\right) =0\, .
\end{equation}
Equation (\ref{key.eq}) was first introduced as the {\it key equation} in \cite{GarciaDiaz2022b} and it imposes nontrivial constraints on the functions $A(r)$, $B(\theta)$, $X(r)$, and $Y(\theta)$ appearing in the general separation ansatz (\ref{GS.At}) and (\ref{GS.Pt}). The electromagnetic potentials in a Kerr-like spacetime must satisfy this fundamental relation.

\subsection{\texorpdfstring{Electromagnetic potentials generating the Kerr-Newman-NUT-$\Lambda$ NLED generalizations}{} }

The general form of the electromagnetic potentials $A_t$ and $ \tilde{\mathcal{P}}_t$, as defined by the arbitrary functions $A(r)$, $B(\theta)$, $X(r)$, and $Y(\theta)$ in Eqs.~(\ref{GS.At}) and (\ref{GS.Pt}), suggests that numerous nonlinear electrodynamics solutions generalizing the Kerr--Newman--NUT spacetime could be constructed by employing higher-degree polynomial expansions in $r$ and $\cos\theta$. Motivated by this idea, we adopt the following Ansatz:
\begin{equation} \label{eq:ansatz}
    \begin{split}
        X(r) & = -Q_e r + \sum_{t}^{} C_{t} r^t,\quad
        Y (\theta) = Q_m \,(a \cos{\theta}+n) + \sum_{s}^{} D_{s} (a
        \cos{\theta}+n)^s,\\
        A(r) & = Q_m \, r + \sum_{l}^{} H_{l} r^l,\quad
        B(\theta) =  Q_e \,(a \cos{\theta}+n) +  \sum_{k}^{} G_{k} (a \cos{\theta}+n)^k,
    \end{split}
\end{equation}
where $C_{t}$, $D_{s}$, $H_{l}$, and $G_{k}$ are constants subject to the constraint imposed by the key equation (\ref{key.eq}). The leading terms in these expansions reproduce the Kerr-Newman-NUT solution. Therefore, the electromagnetic potentials in Eqs. (\ref{GS.At})-(\ref{GS.Pt}) are proposed as
\begin{equation}
A_t = A_{t}^{KN} + \frac{1}{\Sigma} \left( \sum_{t}^{} C_{t} r^t + \sum_{s}^{} D_{s} (a
        \cos{\theta}+n)^s \right), \quad 
\tilde{\mathcal{P}}_{t}= \tilde{\mathcal{P}}_{t}^{KN} + \frac{1}{\Sigma} \left( \sum_{l}^{} H_{l} r^l + \sum_{k}^{} G_{k} (a \cos{\theta}+n)^k \right),
\end{equation}
where the Kerr-Newman-NUT potentials $A_{t}^{KN}$ and $\tilde{\mathcal{P}}_{t}^{KN}$ can be read from Eqs. (\ref{Amu.KN.NUT})-(\ref{pmu.KN.NUT}). These are the potentials we introduce into the Lagrangian integrability condition (\ref{key.eq}) to determine the constants $C_{t}$, $D_{s}$, $H_{l}$, and $G_{k}$.
Within the polynomial aligned ansatz considered here for the Kerr-Newman-NUT-$\Lambda$ spacetime, the key equation (\ref{key.eq}) selects two admissible cases. These are specified by the following sets of coefficients:

{\bf Case 1 (Cubic Solution):}

This solution is characterized by the nonlinearity parameter $\beta$, where $X(r)$ and $A(r)$ are  cubic polynomials in $r$, while $Y(\theta)$ and $B(\theta)$ are cubic in $(a \cos{\theta}+n)$.  The non-zero coefficients in the expansions (\ref{eq:ansatz}) are

\begin{align}
C_1 &= -\beta Q_e n^2, & C_2 &= -2n\beta Q_m(1+\beta n^2), & C_3 &= \beta Q_e, \nonumber\\
D_1 &= \beta Q_m n^2(2 + \beta n^2), & D_2 &= -2n\beta Q_m(1+\beta n^2), &D_3 &= \beta Q_m (1+\beta n^2), \nonumber\\
H_1 &= \beta Q_m n^2, & H_2 &= -2n\beta Q_e(1+\beta n^2), & H_3 &= -\beta Q_m, \nonumber\\
\label{coef1} G_1 &= \beta Q_e n^2(2 + \beta n^2),&G_2 &= -2n\beta Q_e(1+\beta n^2), &G_3 &= \beta Q_e (1+\beta n^2).
\end{align}

{\bf Case 2 (Quartic Solution):}
This second solution is characterized by the nonlinearity parameter $\xi$, where $A(r)$, $B(\theta)$, $X(r)$, and $Y(\theta)$ are polynomials of fourth degree. The non-zero coefficients for this solution in the  expansion (\ref{eq:ansatz}) are

\begin{equation}\label{Coef2}
    \begin{split}
      &    \quad C_2 = \frac{3\xi Q_e n^2}{4}, \quad C_4 = \frac{\xi Q_e}{4},\\
D_1 &= -\frac{n^3 \xi Q_e}{2}, \quad D_2 = \frac{3n^2 \xi Q_e}{4}, \quad D_4 = -\frac{\xi Q_e}{4}, \\
 & \quad H_2 = -\frac{3\xi Q_m n^2}{4}, \quad H_4 = -\frac{\xi Q_m}{4}, \\
G_1 &= \frac{n^3 \xi Q_m}{2}, \quad G_2 = -\frac{3n^2 \xi Q_m}{4}, \quad G_4 = \frac{\xi Q_m}{4}.
    \end{split}
\end{equation}
These solutions describe black holes that have inner, outer, and cosmological horizons, depending on the particular values of the
angular momentum $a$, mass $m$, cosmological constant $\Lambda$, NUT parameter $n$, electric and magnetic charges $Q_e, Q_m$, and the nonlinear parameters  $\beta$ or $\xi$ .

One possible explanation of the existence of only two NLED Kerr-like generalizations is that by assuming the form of the electromagnetic potentials as
in Eqs. (\ref{GS.At})-(\ref{GS.Pt})
\begin{equation}
A_t =\frac{X(r)+Y(\theta)}{\Sigma}, \quad \tilde{\mathcal{P}}_t =\frac{A(r)+B(\theta)}{\Sigma}, \nonumber
\end{equation}
[because that is the solution of the Maxwell-Faraday equations under the alignment conditions in the form of Eqs. (\ref{AC.At.r})], particularly the denominator as $\Sigma= r^2 + (n+a\cos {\theta})^2$,
i.e. a quadratic polynomial, we restrict the electromagnetic multipolar structure to be a dipolar one, in which the electric field,
\begin{equation}
E_{r}= \partial_{r}A_{t}= \frac{X^{\prime}(r)}{\Sigma}- \frac{\Sigma^{\prime}(X(r)+Y(\theta))}{\Sigma^2} \approx
\frac{X^{\prime}(r)}{r^2}- \frac{[X(r)+Y(\theta)]}{r^3}.
\end{equation}
This restriction, through the Einstein equations, leads to a metric function $\Delta_{r}$ that is a fourth order polynomial at most;
therefore through our assumption we are allowing at most an electromagnetic dipolar structure. The guess would be that to generate a solution with a more complex electromagnetic multipolar structure, we should free or extend the denominator assumption of a quadratic polynomial, i.e. to propose electromagnetic potentials of the form $A_{t} \sim Y_{lm}(\theta \phi) /r^{l+1}$ with $l > 1$. Moreover, the form of the electromagnetic potentials Eqs. (\ref{GS.At})-(\ref{GS.Pt}), implies the vanishing of the $\phi$ components of the electric and magnetic fields, $E_{\phi}=0= B_{\phi}$. In a spherical-harmonic language this corresponds to the azimuthal harmonic number $m_{\rm az}=0$, not to be confused with the black-hole mass $m$.

Although exhaustive numerical and algebraic exploration did not yield higher-degree polynomial solutions, a formal proof of uniqueness is beyond the scope of this work.

In the following sections we will analyze each of these two NLED generalizations in detail, focusing on their horizon structure and the conditions required for degenerate horizons.
\subsection{\texorpdfstring{Einstein Field Equations: Differential Elementary Equation}{}}

Taking into account that the components of the tensors appearing in Einstein's field equations satisfy the following relations,
\begin{equation}
    \frac{T_{rr}}{g_{rr}}+\frac{T_{\theta\theta}}{g_{\theta\theta}}=\frac{\tensor{T}{^\alpha_\alpha}}{2},
\end{equation}
we obtain the following linear relations among the components of the energy--momentum tensor
\begin{equation}
    \begin{pmatrix}
        \dfrac{1}{\Sigma^2} & \dfrac{a^2}{\Sigma^2} & -\dfrac{2a}{\Sigma^2} \\
        -\dfrac{\chi}{\Sigma^2} & -\dfrac{a(\Sigma+a\chi)}{\Sigma^2} & \dfrac{\Sigma+2a\chi}{\Sigma^2} \\
        \dfrac{\chi^2}{\Sigma^2} & \dfrac{(\Sigma+a\chi)^2}{\Sigma^2} & -\dfrac{2\chi(\Sigma+a\chi)}{\Sigma^2}
    \end{pmatrix}
    \begin{pmatrix}
        -\dfrac{\Delta_r^2\, T_{rr}}{\Xi^2} \\
        \dfrac{\Delta_\theta^2 \sin^2\theta\, T_{\theta\theta}}{\Xi^2} \\
        0
    \end{pmatrix}
    =
    \begin{pmatrix}
        T_{tt} \\
        T_{t\phi} \\
        T_{\phi\phi}
    \end{pmatrix},
\end{equation}
which can be inverted to yield
\begin{equation}
    \begin{pmatrix}
        (\Sigma+a\chi)^2 & 2a(\Sigma+a\chi) & a^2 \\
        \chi^2 & 2\chi & 1 \\ 
        \chi(\Sigma+a\chi) & \Sigma+2a\chi & a
    \end{pmatrix}
    \begin{pmatrix}
        T_{tt} \\
        T_{t\phi} \\
        T_{\phi\phi}
    \end{pmatrix}
    =
    \begin{pmatrix}
        -\dfrac{\Delta_r^2\, T_{rr}}{\Xi^2} \\
        \dfrac{\Delta_\theta^2 \sin^2\theta\, T_{\theta\theta}}{\Xi^2} \\
        0
    \end{pmatrix}.
\end{equation}
Previous relationships indicate that the only components of Einstein's equations that give us information will be the components with indices $rr$ and $\theta\theta$, since the rest of the equations will be a linear combination of the others.
The equation to be solved in order to determine the function $f(r)$ is $G_{rr}+\Lambda g_{rr}=8\pi T_{rr}$, and that equation reduces to

\begin{equation}\label{diff.eq.fr}
    ED+BH-\frac{Q_e^2+Q_m^2}{\Sigma^2}
=
\frac{\Sigma f''(r)-4r f'(r)+4f(r)}{4\Sigma^2}.
\end{equation}

In the linear case, the right-hand side of the above equation vanishes.

\subsection{Logical structure of the proof and explicit verification}
\label{subsec:proof-structure}

For the convenience of the reader, we summarize here the logical structure that proves that the line elements and the aligned potentials presented below solve the field equations.

\paragraph*{Step 1: Alignment solves the differential Maxwell sector.}
The aligned ansatz implies that both $F_{\mu\nu}=dA$ and ${}^{\star}P_{\mu\nu}=d\tilde{\mathcal P}$ are exact two-forms generated by the potentials in Eqs.~(\ref{GS.At})--(\ref{GS.Pt}). Therefore the Bianchi identities and the source-free Maxwell equations are reduced to the compatibility of the aligned potentials with the constitutive relations. In other words, alignment does not merely simplify the tensor structure: it makes the electromagnetic sector compatible with stationarity, axial symmetry, and the principal directions of the geometry.

\paragraph*{Step 2: The key equation guarantees constitutive integrability.}
The key equation is the local integrability condition for the on-shell Lagrangian. It is therefore a necessary condition for the existence of a NLED description along the aligned solution branch; the question of a global single-valued expression $\mathcal{L}(F,G)$ is separate and, for the rotating families, remains open.

\paragraph*{Step 3: Einstein's equations reduce to a single radial equation.}
For the Kerr--Newman--NUT--$\Lambda$-type metric with $\Delta_r=\Delta_r^{KN}+f(r)$, the algebraic relations among the components of $T^{\mu}{}_{\nu}$ imply that the independent content of Einstein's equations is carried by the $rr$ and $\theta\theta$ components. Equation~(\ref{diff.eq.fr}) is therefore the master equation that determines the nonlinear deformation $f(r)$ once the aligned fields $(E,B,D,H)$ are known.

\paragraph*{Step 4: Componentwise verification.}
In the linear case, Eq.~(\ref{diff.eq.fr}) gives $f(r)=0$, so the standard Kerr--Newman--NUT--$\Lambda$ solution is recovered. In the cubic and quartic cases, substitution of the corresponding aligned fields into Eq.~(\ref{diff.eq.fr}) yields the explicit deformations displayed in Sections~\ref{CubicSol} and \ref{QuarticSol}. The complete componentwise checks were independently carried out in Maple: the alignment identities, the key equation, and the independent Einstein equations simplify identically to zero for the line elements and potentials reported in this work. This confirms that the presented metrics are exact solutions of the Einstein--Maxwell system in the linear case and of the Einstein--NLED system in the cubic and quartic cases.

\paragraph*{Conservation versus trace.}
A nonvanishing trace $T^{\mu}{}_{\mu}\neq 0$ is expected in nonlinear electrodynamics and signals the breaking of conformal invariance. This does not conflict with consistency of the matter sector: because the action is diffeomorphism invariant and the generalized Maxwell equations hold, the stress tensor remains covariantly conserved on shell, $\nabla_\mu T^{\mu\nu}=0$, in all three cases considered here.

\section{\texorpdfstring{Cubic NLED  generalizations of the Kerr-Newman-NUT-$\Lambda$  metric: The cubic vector potential}{} }\label{CubicSol}

The cubic NLED generalization, called the Kerr–Newman–NUT–$\Lambda$-type spacetime with a cubic vector potential is, in Boyer–Lindquist coordinates,
\begin{equation}\label{m.POT3}
\begin{split}
ds^2&=\frac{\Sigma}{\Delta_{r}}\,{ dr}^2-{\frac{\Delta_{r}}{\Sigma \Xi^2}}\left( {{dt}} -
\chi{ d \phi} \right)^2+\frac{\Sigma}{\Delta_{\theta}}\,{ d \theta}^2+ \frac{\sin^2{\theta} \Delta_{\theta}}{\Sigma \Xi^2}  \left[ {a {dt}}
-\left(\Sigma+a\chi\right) { d \phi} \right]^2,\\
\Delta_{r} &:=r^2-2mr+(1-n^2\Lambda)(a^2-n^2)-\frac{\Lambda r^2}{3}(r^2+a^2+6n^2)+ ({Q_e}^2+{Q_m}^2)[1+\beta(n^2-r^2)]^2(1+\beta n^2),\\
 \Delta_{\theta} &= 1 + \frac{\Lambda}{3}a \cos \theta\left(a\cos\theta+4n\right),\quad
 \Sigma={{r^2+(a\cos\theta+n)^2}};
 \quad
 \chi:=a \sin^2 \theta+2n(1-\cos\theta);\quad\Xi=1+\frac{\Lambda}{3}a^2;
\end{split}
\end{equation}
besides the usual Kerr-Newman-NUT-$\Lambda$ parameters, an additional nonlinearity parameter $\beta$ is introduced through the deformation function $f(r)$, 
\begin{equation}
\begin{split}
    \Delta_{r}&={\Delta}_{r}^{KN}+ f(r)\\ 
    &={\Delta}_{r}^{KN}+ ({Q_e}^2+{Q_m}^2)   \left\{ [1+\beta(n^2-r^2)]^2(1+\beta n^2)-1 \right\},
\end{split}
\end{equation}
where ${\Delta}_{r}^{KN}$ corresponds to the linear case, the KN-NUT  solution of the Einstein-Maxwell equations given in Eq. (\ref{metric1}), and the metric function $f(r)$ is determined from the Einstein-NLED differential equation

\begin{equation}
\frac{\Sigma\, f''(r)- 4 r f'(r)+ 4 f(r)}{4 \Sigma^{2}}=-\beta \frac{(Q_e^2 + Q_m^2)}{\Sigma^{2}}\Big[(1+\beta n^2)(n+a\cos\theta)^2\big(1+\beta n^2-3\beta r^2\big)-(1+\beta n^2)^2 r^2- n^2(\beta^2 n^4+3\beta n^2+3)\Big],
\end{equation}

whose solution is

\begin{equation}\label{f.cubic}
     f(r)_{cubic}=({Q_e}^2+{Q_m}^2)   \left\{ [1+\beta(n^2-r^2)]^2(1+\beta n^2)-1 \right\}.
\end{equation}

 The electromagnetic potentials of the Cubic NLED-KN-NUT-$\Lambda$ solution, the cubic vector potentials, are given by (\ref{GS.At})-(\ref{GS.Pt}) where the functions $X(r)$, $Y(\theta)$, $A(r)$ and $B(\theta)$ are given by (\ref{eq:ansatz}) with coefficients in (\ref{coef1})
\begin{equation}\label{Amu.KN.NUT.NLE3}
\begin{split}
    A = & \left\{{\frac{-Q_e r[1+\beta(n^2-r^2)]-2n\beta Q_m  r^2(1+\beta n^2)}{\Sigma \Xi}}\right\}\left( {{dt}} -\chi{ d \phi} \right)\\ & +\frac{Q_m (a\cos {\theta}+n)\left(1+\beta a^2 \cos^2 \theta\right)(1+\beta n^2)}{a\Sigma \Xi} \left[ {a {dt}}-\left(\Sigma+a\chi\right) { d \phi} \right]
\end{split}
\end{equation}
and
\begin{equation}\label{pmu.KN.NUT.NLE3}
\begin{split}
    \tilde{\mathcal{P}} = & \left\{{\frac{Q_m r[1+\beta(n^2-r^2)]-2n\beta Q_e  r^2(1+\beta n^2)}{\Sigma \Xi}}\right\}\left( {{dt}} -\chi{ d \phi} \right)\\&+\frac{Q_e (a\cos {\theta}+n)\left(1+\beta a^2 \cos^2 \theta\right)(1+\beta n^2)}{a\Sigma \Xi}\left[ {a {dt}}-\left(\Sigma+a\chi\right) { d \phi} \right].
\end{split}
\end{equation}
The electromagnetic field components are given by

\begin{align}
   \label{E.POT3} E&=-Q_e\left[\frac{\left(r^2-x^2\right)(1+\beta n^2)+\beta r^2\left(r^2+3x^2\right)}{\left(r^2+x^2\right)^2}\right]
    +2Q_m \frac{rx(1+\beta n^2)\left(1+\beta n^2+\beta x^2\right)}{\left(r^2+x^2\right)^2},\\
    \label{B.POT3} B&=Q_m(1+\beta n^2)\left[\frac{\left(r^2-x^2\right)(1+\beta n^2)+\beta x^2\left(3r^2+x^2\right)}{\left(r^2+x^2\right)^2}\right]+2Q_e\frac{rx\left(1+\beta n^2-\beta r^2\right)}{\left(r^2+x^2\right)^2},\\
   \label{D.POT3} D&=-Q_e(1+\beta n^2)\left[\frac{\left(r^2-x^2\right)(1+\beta n^2)+\beta x^2\left(3r^2+x^2\right)}{\left(r^2+x^2\right)^2}\right]+2Q_m \frac{rx\left(1+\beta n^2-\beta r^2\right)}{\left(r^2+x^2\right)^2},\\
  \label{H.POT3}  H&=Q_m\left[\frac{\left(r^2-x^2\right)(1+\beta n^2)+\beta r^2\left(r^2+3x^2\right)}{\left(r^2+x^2\right)^2}\right]
    +2Q_e \frac{rx(1+\beta n^2)\left(1+\beta n^2+\beta x^2\right)}{\left(r^2+x^2\right)^2},
\end{align}  
where $x=n+a\cos\theta$. For $\beta =0$ we recover the Kerr-Newman-NUT-$\Lambda$ solution. The case $n=0$ was presented in \cite{GarciaDiaz2022b}.
The trace of the stress-energy tensor is nonvanishing, due to the nonlinear electromagnetic field, characterized by the parameter $\beta$,
\begin{equation}
\label{T.POT3} \pi   {T}{^\mu_\mu}=-\frac{ \beta (Q_e^2+Q_m^2)\left(1+\beta n^2-3\beta r^2\right)(1+\beta n^2)}{2\left[r^2+(a\cos\theta+n)^2\right]} \, .
\end{equation}
with the corresponding Lagrangian is
\begin{equation}\label{Lag.EDBH}
\mathcal{L}=\frac{ED-BH}{2}+\pi \tensor{T}{^\mu_\mu}. 
\end{equation}

The Weyl conformal invariance is violated through the introduction of the NLED parameter $\beta$; for $\beta = 0$ the traceless energy momentum tensor ${T}{^\mu_\mu}=0$ is recovered. The corresponding Lagrangian is determined by substituting Eqs. (\ref{E.POT3})-(\ref{T.POT3}) in Eq. (\ref{Lag.EDBH}).

The NLED generalization for the Kerr metric with the {\it cubic vector potential} was presented in \cite{GarciaDiaz2022b} and represents a rotating charged BH characterized by the usual parameters and a NLED nonlinearity parameter $\beta$. Its asymptotics can be de Sitter or anti-de Sitter, and even flatness, depending on $\beta$ and $\Lambda$. This BH can possess one, two or three horizons; the latter is a cosmological horizon in the de Sitter case. In case the BH is static ($a=0$), the solution corresponds to a NLED generalization of the Reissner-Nordstrom-NUT-$\Lambda$ solution.

The metric (\ref{m.POT3}) comprises at least 24 spacetimes that are shown in Table \ref{tab:spacetimes}. Particularly there are 8 nonlinear electromagnetic cases.

\subsection{Singularities of the Cubic NLED-Kerr-Newman-NUT spacetime}\label{VA}
A complete analysis of the singularities of the Kerr-Newman-NUT-$\Lambda$ metric can be carried through according to \cite{Grenzebach2014}. Here we just comment on the modifications introduced by the nonlinearity parameter $\beta$. The metric (\ref{m.POT3}) becomes singular if (a) $\Sigma=0$, (b) $\Delta_{r}=0$, (c) $\Delta_{\theta}=0$ and (d) $\sin\theta =0$.
For the cases (a), (c) and (d) the analysis presented in \cite{Grenzebach2014} can be taken over. In case (b), however, the parameter $\beta$ modifies the horizon structure. In what follows we briefly comment on these singularities.

\begin{itemize}
\item[{(a)}]
If $\Sigma=0$, a ring singularity exists if both conditions are met: $r=0$ and $\cos\theta=-\frac{n}{a}$. If $n^2 >a^2$, there is no ring singularity, since $\Sigma$ is non-zero everywhere and the entire sphere $r=0$ is regular. This singularity also appears in the explicit Kretschmann scalar for the cubic family and its asymptotic values found in Appendix ~\ref{KretS}.
\item[{(b)}] The function $\Delta_{r}$ in this case is a fourth degree polynomial with real coefficients; therefore there are 4, 2 or zero real roots.
We analyze in detail in the next subsection the horizons that arise when $\Delta_{r}=0$.
\item[{(c)}] The singularity arising from $\Delta_{\theta}=0$, rather than being on a sphere $r=$constant, it is situated on a cone $\theta=$ constant, defined by
\begin{equation}
a\cos\theta_\pm=\frac{4}{3}\left(-\Lambda n\pm\sqrt{\Lambda\left(\Lambda n^2-\frac{3}{4}\right)}\right);
\end{equation}
such that restricting to $4 n^2 \Lambda < 3$ we can be sure that $\Delta_{\theta}$ has no zeros.
\item[{(d)}] When using spherical polar coordinates there is always a singularity on the axis $\sin \theta=0$; this is a true singularity. For details in managing
on which part of the axis (half-axis $\theta=0$ or $\theta= \pi$) the singularity is situated can be consulted \cite{Manko2005}.
\end{itemize}

\subsection{\texorpdfstring{Horizons of the Cubic NLED-Kerr-Newman-NUT-$\Lambda$ spacetime.}{}}

As was stated previously, we shall identify the event horizon with the largest positive root of $\Delta_r=0$, which we call $r_+$, corresponding to the outermost null hypersurface bounding the black hole region. The effective cosmological horizon, where the curvature invariants are finite, is said to be an apparent horizon.

The determination of the horizons amounts to analyzing the roots of
\begin{equation}\label{Delta.POT3}
\begin{split}
\Delta_{r}& = \left[(Q_e^2+Q_m^2)(1+\beta n^2)\beta^2-\frac{\Lambda}{3}\right]r^4+\left[1-\frac{\Lambda}{3}(a^2+6n^2)-2\beta(1+\beta n^2)^2(Q_e^2+Q_m^2)\right]r^2\\
&-2mr+(1-\Lambda n^2)(a^2-n^2)+(Q_e^2+Q_m^2)(1+\beta n^2)^3=0,
\end{split}
\end{equation}
whose solutions are given by (\ref{rc12}) and (\ref{rc34}). Note that there is no term cubic in $r$.

We denote the roots as cosmological horizon $(r_c)$, outer horizon $(r_+)$, inner horizon $(r_-)$ and the fourth root as
$(r_0)$, in such a way that $r_c > r_{+} > r_{-} > r_0$. Then $\Delta_r$ in the metric (\ref{m.POT3}) can be written as
\begin{equation}\label{m.Pot3.Dark}
     \Delta_r= \left[(Q_e^2+Q_m^2)(1+\beta n^2)\beta^2-\frac{\Lambda}{3}\right](r-r_0)(r-r_-)(r-r_+)(r-r_c).
\end{equation}

To simplify the analysis, we used the Cardano–Vieta relations (relations between roots and coefficients of a polynomial). We note that, in this case, for the polynomial in Eq.~(\ref{Delta.POT3}), the following relations hold:

\begin{equation}\label{POT3.Cardano.Vieta}
\begin{split}
r_0 &= -(r_- + r_+ + r_c), \\
(Q_e^2 + Q_m^2)(1 + \beta n^2)\beta^2 - \frac{\Lambda}{3} &= -\frac{2m}{(r_+ + r_-)(r_+ + r_c)(r_- + r_c)}, \\
1 - \frac{\Lambda}{3}(a^2 + 6n^2) - 2\beta(1 + \beta n^2)^2(Q_e^2 + Q_m^2) &= 2m \left[ \frac{r_+^2 + r_-^2 + r_c^2 + r_+ r_- + r_+ r_c + r_- r_c}{(r_+ + r_-)(r_+ + r_c)(r_- + r_c)} \right], \\
(1 - \Lambda n^2)(a^2 - n^2) + (Q_e^2 + Q_m^2)(1 + \beta n^2)^3 &= 2m \left[ \frac{(r_- + r_+ + r_c) r_- r_+ r_c}{(r_+ + r_-)(r_+ + r_c)(r_- + r_c)} \right],
\end{split}
\end{equation}
provided both $\beta \ne 0$ and $\Lambda \ne 0$.
Since the general case is extremely complicated, we analyze particular cases of degenerate horizons or
multiple roots in order to constrain the allowed ranges of the black hole (BH) parameters.

\begin{figure}[htbp]
\centering
\subfloat[$a=0,~n=0$, $\beta=0.11/m^2$.]{\label{Pot3.f11}\includegraphics[width=0.40\textwidth]{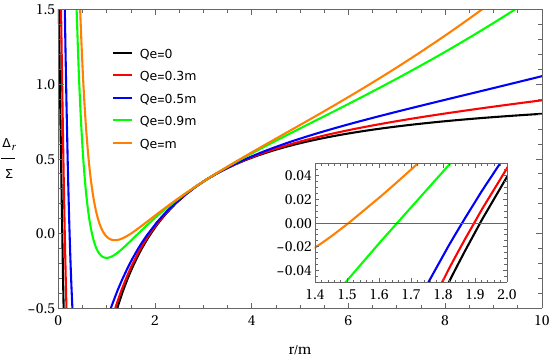}}
\hspace{0.5cm} 
\subfloat[$a=0,~n=0$, $Q_e=0.5m$.]{\label{Pot3.f21}\includegraphics[width=0.40\textwidth]{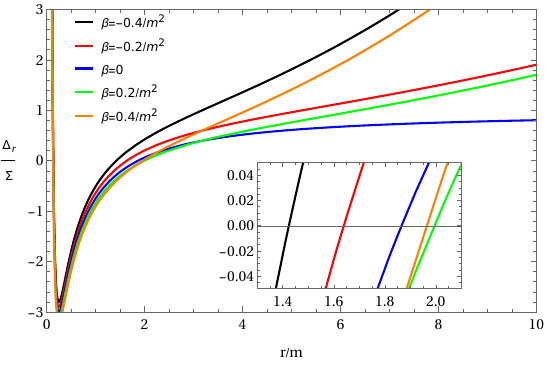}}
\\[-2ex] 

\subfloat[$a=0,~n=0.3m$, $\beta=0.11/m^2$.]{\label{Pot3.f12}\includegraphics[width=0.40\textwidth]{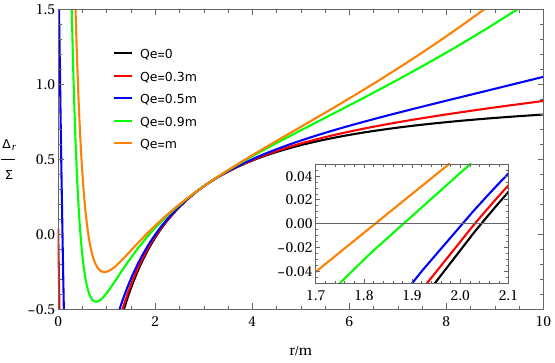}}
\hspace{0.5cm}
\subfloat[$a=0,~n=0.3m$, $Q_e=0.5m$.]{\label{Pot3.f22}\includegraphics[width=0.40\textwidth]{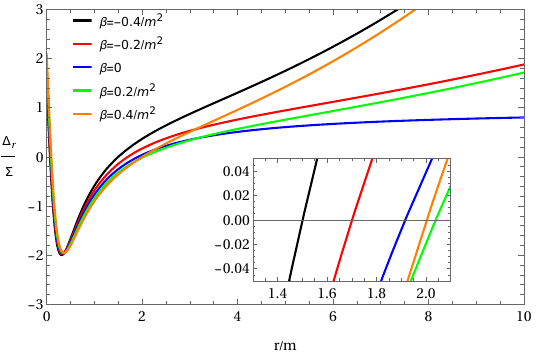}}
\\[-2ex]

\subfloat[$a=0.5m,~n=0$, $\beta=0.11/m^2$.]{\label{Pot3.f13}\includegraphics[width=0.40\textwidth]{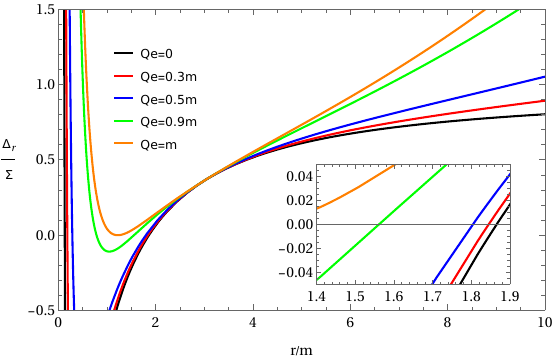}}
\hspace{0.5cm}
\subfloat[$a=0.5m,~n=0$, $Q_e=0.5m$.]{\label{Pot3.f23}\includegraphics[width=0.40\textwidth]{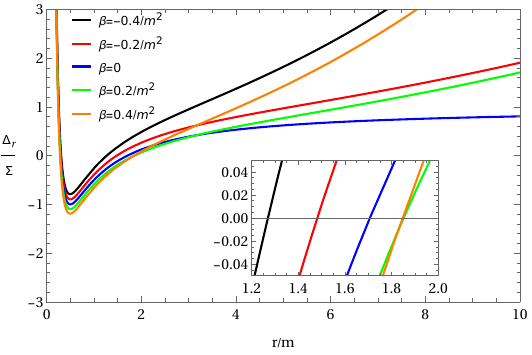}}
\\[-2ex]

\subfloat[$a=0.5m,~n=0.3m$, $\beta=0.11/m^2$.]{\label{Pot3.f14}\includegraphics[width=0.40\textwidth]{image/Pot3.m1.a0.5.n0.3.beta0.11.Lambda0.g0Q0.0.3.0.5.09.1.pdf}}
\hspace{0.5cm}
\subfloat[$a=0.5m,~n=0.3m$, $Q_e=0.5m$.]{\label{Pot3.f24}\includegraphics[width=0.40\textwidth]{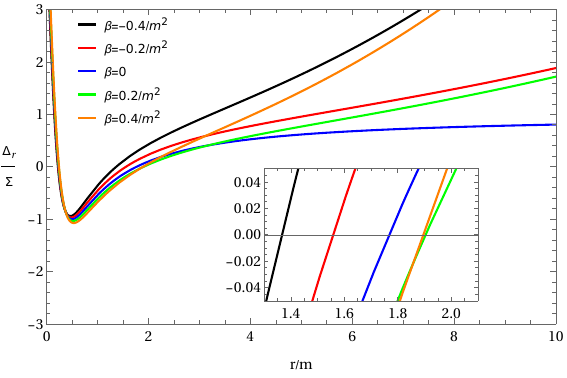}}

\caption{Comparison of the metric function $\Delta_{r}(r)/ \Sigma$ for different values of $a$, $n$, $\beta$ and $Q_e$. The first column corresponds to fixed $\beta=0.11/m^2$ while $Q_e$ varies, whereas in the second column the charge is fixed at $Q_e=0.5m$ and $\beta$ varies. Each row corresponds to different values of $(a,n)$: (a,b) $a=0$, $n=0$; (c,d) $a=0$, $n=0.3m$; (e,f) $a=0.5m$, $n=0$; and (g,h) $a=0.5m$, $n=0.3m$.}
\label{Delta.beta.Pot3}
\end{figure}

Figs. \ref{Pot3.f11} - \ref{Pot3.f24} show the metric functions $\Delta_{r}/ \Sigma$ as a function of $r$, for $\Lambda=0$.

For fixed BH charge and varying NLED parameter $\beta$ the horizons depend also on the sign of $\beta$. A parameter $\beta > 0.25$ reverses the general tendency of growing the horizon radius as $\beta$ increases; it may be due to the effect of $\beta$ mimicking a positive cosmological constant, presenting the solution a non-flat asymptotics.
For $\beta=0.25$ the horizon radius is maximum $r_{+}=2m$ and does not depend on the rest of the BH parameters.

In \ref{Pot3.f11} - \ref{Pot3.f21} is shown $\Delta_{r}/ \Sigma$ for the RN-NLED generalization, with $a=0$ and $n=0$.
In \ref{Pot3.f11} by fixing $\beta= 0.11/m^2$ for $Q_e = 0; 0.3m; 0.5m; 0.9m; m$; by increasing $Q_e$ the horizon radius decreases. In \ref{Pot3.f21} with fixed $Q_e= 0.5m$ the NLED parameter $\beta$ is varied ; as $\beta$ increases the horizon radius consistently increases, except for $\beta = 0.4m$ where we observe a different tendency.

In \ref{Pot3.f12} - \ref{Pot3.f22} $\Delta_{r}(r)/ \Sigma$ are shown for the static case ($a=0$) with NUT parameter ($n= 0.3m$), i.e. the RN-NUT-NLED generalization; In \ref{Pot3.f12} by fixing $\beta=0.11/m^2$, the BH charge increases, $Q_e= 0; 0.3m; 0.5m; 0.9m; m$, that makes the horizon radius consistently diminishing. In \ref{Pot3.f22} with fixed $Q_e= 0.5m$ is varied the NLED parameter $\beta$; as $\beta$ increases the horizon radius consistently increases, except for $\beta = 0.4m$ where we observe a different tendency.

In Figs. \ref{Pot3.f13} - \ref{Pot3.f24} the metric functions $\Delta_{r}(r)/ \Sigma $ for the rotating NLED Kerr-like generalizations are shown, with fixed angular momentum $a=0.5m$.
In \ref{Pot3.f13} - \ref{Pot3.f23} is shown $\Delta_{r}(r)/ \Sigma$ of the Kerr-Newman NLED generalization, with vanishing NUT parameter ($n=0$); in \ref{Pot3.f13} with fixed NLED parameter $\beta=0.11/m^2$, the BH charge increases, $Q_e= 0; 0.3m; 0.5m; 0.9m; m$, that makes the horizon radius to decrease consistently. In \ref{Pot3.f23}
with fixed $Q_e= 0.5m$ is varied the NLED parameter $\beta$; as $\beta$ increases the horizon radius consistently increases, except for $\beta = 0.4m$ where we observe a different tendency.

In \ref{Pot3.f14} - \ref{Pot3.f24} the metric function $\Delta_{r}(r)/ \Sigma$ for Kerr-Newman-NUT NLED generalization, with fixed NUT parameter $n=0.3$; in \ref{Pot3.f14} with fixed NLED parameter $\beta=0.11/m^2$, the BH charge increases, $Q_e= 0; 0.3m; 0.5m; 0.9m; m$, that makes the horizon radius to decrease consistently. In \ref{Pot3.f24}
with fixed $Q_e= 0.5m$ is varied the NLED parameter $\beta= -0.4/m^2; -0.2/m^2; 0.; 0.2/m^2; 0.4/m^2$; as $\beta$ increases the radius of the horizon consistently increases, except for $\beta = 0.4m$ where we observe the horizon is small.
 
\subsection{Cubic NLED-Reissner-Nordstrom generalization}
 
In the case where $\Lambda = 0$, $n = 0$, and $a = 0$, we obtain the static NLED Reissner-Nordstr\"om (RN) generalization with both electric and magnetic charges, where $Q^2= Q_m^2 +Q_e^2$. The corresponding metric function $\Delta_{r}$ is given by:
\begin{equation}
    \Delta_{r}= r^2-2mr+ Q^2 (1-\beta r^2)^2\, .
\label{RNmetric}
\end{equation}
The event horizons are located at the positive real roots of the equation $\Delta_{r}(r)=0$. 

In the linear limit ($\beta = 0$), this reduces to the standard static solution of the Einstein-Maxwell equations. If $0 \leq Q^2 < m^2$, it possesses two distinct horizons, $r_{\pm} = m \pm \sqrt{m^2 - Q^2}$. The extremal case occurs when $|Q| = m$, leading to a single degenerate horizon at $r_+ = m$. 

For the non-linear RN generalizations ($\beta \neq 0$), the parameter space becomes richer, leading to the following physically distinct cases:
  
\begin{enumerate}
    \item If $|Q| = m$ and $\beta > 0$, the non-linear effects ensure the existence of two distinct horizons bounded by $0 \leq r_- \leq r_+ \leq 2m$.
    
    \item If $|Q| = m$ and $\beta < 0$, the metric function has no real roots, indicating the presence of a naked singularity.
    
    \item For any given non-linear coupling $\beta > 0$, there exists a critical extremal charge $Q_{ext}(\beta) > m$ that yields a degenerate horizon. For instance, if $\beta = \frac{9}{50 m^2}$, the extremal charge is exactly $Q_{ext} = \frac{2\sqrt{5}}{3}m$, which results in a single degenerate event horizon located at $r_+ = \frac{5}{3}m$.
    
    \item If $|Q| < Q_{ext}(\beta)$ for a given positive $\beta$, the metric function always admits two distinct real roots, corresponding to an inner and an outer horizon.
    
    \item If $|Q| > Q_{ext}(\beta)$ for a given positive $\beta$, no horizons exist, and the geometry describes a naked singularity.
\end{enumerate}

\subsection{Effective cosmological contribution of the cubic nonlinear sector}

From Eq. (\ref{Delta.POT3}) for $\Delta_{r}$ it is clear that the NLED parameter $\beta$ can be tuned to compensate the cosmological constant $\Lambda$. Equivalently, NLED leads to a dark energy contribution.
In other words, this case corresponds to a specific value of the cosmological constant
\begin{equation}
\Lambda = 3(Q_e^2 + Q_m^2)(1 + \beta n^2)\beta^2.
\label{Lde}
\end{equation}
that exactly cancels out the contribution of the NLED parameter $\beta$, resulting in a reduction in the degree of $\Delta_{r}$.
In this case Eq. (\ref{Delta.POT3}) simplifies to a quadratic polynomial, $\Delta_r = a_2 r^2 + a_1 r + a_0$, with coefficients
\begin{align}
a_2 &=1-xz, &a_1&= -2m, &a_0&=xy+w, & & \\
x&=(1 + \beta n^2)(Q_e^2 + Q_m^2), &y&=(1 + \beta n^2)^2 - 3\beta^2 n^2 w, &z&=\beta\left[2 + \beta(a^2 + 8n^2)\right], &w&=a^2-n^2.
\end{align}
In this case, the horizons are located at
\begin{equation}
r_{\pm} = \frac{m \pm \sqrt{m^2 - a_2 a_0}}{a_2},
\end{equation}
and they occur when $a_2 a_0 \leq m^2$ and $a_2 > 0$ (ensuring real and positive roots). The extremal condition $m^2 = a_2 a_0$ yields the relations
\begin{equation}
0 \leq Q_e^2 + Q_m^2=x_{\mp}= 2\left(\frac{m^2 - w}{1 + \beta n^2}\right) \left(\frac{1}{y-zw\mp\sqrt{(zw+y)^2-4yzm^2}}\right)\leftrightarrow \Lambda = \frac{6(m^2 - w)\beta^2}{y-zw\mp\sqrt{(zw+y)^2-4yzm^2}}. \label{DM3}
\end{equation}

If $\beta = 0$ (i.e., $\Lambda = 0$), in $x_+$, \eqref{DM3} reduces to the familiar condition,
\begin{equation}
    a^2 - n^2 + Q_e^2 + Q_m^2 \leq m^2.
\end{equation}
The expression $x_-$ is not taken into account because it is indeterminate at $\beta=0$, as $x_+$ reproduces the classical values well, let's see how it behaves in the non-linear case. For the special case $\beta \neq 0$, $n = 0$, and $a = 0$, \eqref{DM3} simplifies to
\begin{equation}
0 \leq Q_e^2 + Q_m^2 = \frac{2m^2}{1 + \sqrt{1 - 8\beta m^2}} \leq 2m^2, \leftrightarrow \Lambda =  \frac{6m^2\beta^2}{1 + \sqrt{1 - 8\beta m^2}}
\end{equation}
which holds for $-\infty < \beta \leq \frac{1}{8m^2}$ and it give us an upper bound $2m^2$ for the charges.
If $\beta \neq 0$, $a \neq 0$, and $n = 0$, the sum of the squared electric and magnetic charges remains positive and well-defined when $\beta \in \left[-\frac{64}{27m^2}, \frac{1}{8m^2}\right]$ and $0 \leq a^2 \leq m^2$.

Starting from equation (\ref{Lde}), it is instructive to explore the limiting behavior of some parameters. For instance, adopting the observational estimate of the cosmological constant, $\Lambda \approx 1.11\times10^{-52}$, and setting the NUT charge to zero, one obtains
\begin{equation}
|Q\beta| = \sqrt{\frac{\Lambda}{3}} \approx 6.08\times10^{-27}.
\end{equation}

Moreover, using the upper bound for the NUT parameter from Solar System observations, $|n| \le 0.032$, as reported in \cite{Hackmann2012}, one can solve equation (\ref{Lde}) for the cubic expression
\begin{equation}
3Q^{2}n^{2}\beta^{3} + 3Q^{2}\beta^{2} - \Lambda = 0
\quad \leftrightarrow \quad
\beta = \frac{2\cos\!\left[\frac{1}{3}\arccos\!\left(\frac{9n^{4}\Lambda - 2Q^{2}}{2Q^{2}}\right) + \frac{2k\pi}{3}\right] - 1}{3n^{2}},
\end{equation}
which implies the consistency condition
\begin{equation}
0 \leq \frac{9n^{4}\Lambda}{2} \leq Q^{2}\leftrightarrow 2.2886 \times 10^{-29}m\leq |Q|.
\end{equation}

These results suggest that, for realistic cosmological values of $\Lambda$ and small NUT charge, the nonlinearity parameter $\beta$ remains extremely small, reinforcing the perturbative nature of the deviation from the Maxwell regime.


\vspace{1em}



\section{\texorpdfstring{Quartic NLED generalizations of the Kerr-Newman-NUT-$\Lambda$  spacetime}{}}\label{QuarticSol}

{For the particular case in which a nonlinear quartic contribution arises in the electromagnetic potentials, the metric function $f(r)$ is determined by the following differential equation}

\begin{equation}
\frac{\Sigma\, f''(r)- 4 r f'(r)+ 4 f(r)}{4 \Sigma^{2}}=\xi r \frac{(Q_e^2 + Q_m^2)}{2\Sigma^{2}}\Big[r^2-3(n+a\cos\theta)^2\Big],
\end{equation}

whose solution is

\begin{equation}\label{f.quartic}
     f(r)_{quartic}=-\xi ({Q_e}^2+{Q_m}^2) r^3.
\end{equation}

The Quartic NLED-Kerr-Newman-NUT-$\Lambda$ solution corresponds to an electromagnetic potential that is quartic in the radial coordinate and in $\cos \theta$; in Boyer-Lindquist coordinates the metric is given by

\begin{equation}\label{m.POT4}
\begin{split}
ds^2&=\frac{\Sigma}{\Delta_{r}}\,{dr}^2-{\frac{\Delta_{r}}{\Sigma \Xi^2}}\left( {{dt}} -
\chi{ d \phi} \right)^2+\frac{\Sigma}{\Delta_{\theta}}\,{ d \theta}^2+ \frac{\sin^2{\theta} \Delta_{\theta}}{\Sigma \Xi^2}  \left[ {a {dt}}
-\left(\Sigma+a \chi \right) { d \phi} \right]^2,\\
\Delta_r &=r^2-2mr+(1-n^2\Lambda)(a^2-n^2)-\frac{\Lambda r^2}{3}(r^2+a^2+6n^2)+(Q_e^2+Q_m^2)(1-\xi r^3),\\
 \Delta_{\theta} &= 1 + \frac{\Lambda}{3}a \cos \theta\left(a\cos\theta+4n\right),\quad
 \Sigma={{r^2+(a\cos\theta+n)^2}};
 \quad
 \chi:=a \sin^2 \theta+2n(1-\cos\theta);\quad\Xi=1+\frac{\Lambda}{3}a^2 \, .
\end{split}
\end{equation}
The parameters are angular momentum $a$, mass $m$, cosmological constant $\Lambda$, NUT parameter $n$, electric and magnetic charge, $Q_e$, $Q_m$, and the NLE parameter, $\xi$, coming from the NLE contribution that has been introduced solving the Einstein equations through the metric function 
\begin{equation}
\Delta_{r}= \Delta^{KN}_{r} - \xi ({Q_e}^2+{Q_m}^2) r^3,    
\end{equation}
where $\Delta^{KN}_{r}$ corresponds to the linear case, the solution of the Einstein-Maxwell equations, the KN-NUT solution, given in Eq. (\ref{metric1}).
The electromagnetic potentials are given by, Eqs. (\ref{GS.At})-(\ref{GS.Pt}) where the functions $X(r)$, $Y(\theta)$, $A(r)$ and $B(\theta)$ are given by (\ref{eq:ansatz}) with coefficients in (\ref{Coef2}),
\begin{align}\label{Amu.KN.NUT.NLE4}
A &= {\frac{-Q_e r\left[1-\frac{\xi r}{4}\left(r^2+3n^2\right)\right]}{\Sigma \Xi}}\left( {{dt}} -\chi{  d \phi} \right) \nonumber \\
& +\frac{(a\cos {\theta}+n)}{a\Sigma \Xi}\left[Q_m-\frac{\xi Q_e a^2 \cos^2 \theta(a\cos\theta+3n)}{4}\right]  [ {a {dt}}-\left(\Sigma+a \chi \right) { d \phi} ],  \\
\tilde{\mathcal{P}} & = {\frac{Q_m r\left[1-\frac{\xi r}{4}\left(r^2+3n^2\right)\right]}{\Sigma \Xi}}\left( {{dt}} -\chi{ d \phi} \right) \nonumber\\
& + \frac{(a\cos {\theta}+n)}{a\Sigma \Xi}\left[Q_e+\frac{\xi Q_m a^2 \cos^2 \theta(a\cos\theta+3n)}{4}\right]  [ {a {dt}}-\left(\Sigma+a\chi\right) { d \phi} ] \, . \label{Pmu.KN.NUT.NLE4}
\end{align}
The corresponding electromagnetic field components are
\begin{align}
 E&=-Q_e\left[\frac{r^2-x^2+\xi n ^3 xr}{\left(r^2+x^2\right)^2}+\frac{\xi r}{2}\right]+\frac{2Q_m xr}{\left(r^2+x^2\right)^2},\label{E.POT4}\\
 B&=Q_m\left[\frac{r^2-x^2}{\left(r^2+x^2\right)^2}\right]+Q_e \left[\frac{2rx-\frac{\xi  n^3}{2}(r^2-x^2)}{\left(r^2+x^2\right)^2}-\frac{\xi x}{2}\right],\label{B.POT4}\\
 D&=-Q_e\left[\frac{r^2-x^2}{\left(r^2+x^2\right)^2}\right]+Q_m \left[\frac{2rx-\frac{\xi  n^3}{2}(r^2-x^2)}{\left(r^2+x^2\right)^2}-x\frac{\xi x}{2}\right],\label{D.POT4}\\
 H&=Q_m\left[\frac{r^2-x^2+\xi n ^3 xr}{\left(r^2+x^2\right)^2}+\frac{\xi r}{2}\right]+\frac{2Q_e xr}{\left(r^2+x^2\right)^2},\label{H.POT4}
\end{align}
where $x=n+a\cos\theta$. The trace of the energy-momentum tensor is
\begin{equation}\label{T.POT4}
\pi {T}{^\mu_\mu}=-\frac{3(Q_e^2+Q_m^2)r\xi}{4\left[r^2+(a\cos\theta+n)^2\right]}.
\end{equation}
The corresponding Lagrangian is determined by substituting Eqs. (\ref{E.POT4})-(\ref{T.POT4}) in Eq. (\ref{Lag.EDBH}). The explicit on-shell Lagrangian for the cubic and quartic family in terms of the invariants in the static limit is presented in Appendix \ref{app:Lagrangian}.

\subsection{\texorpdfstring{Singularities of the Quartic NLED-Kerr-Newman-NUT-$\Lambda$ spacetime}{}}
The metric~\eqref{m.POT4} becomes singular when
$\Sigma=0$ (as well as the explicit Kretschmann scalar for the quartic family and its asymptotic values found in Appendix~\ref{kret.quartic}), $\Delta_r=0$, $\Delta_\theta=0$, or
$\sin\theta=0$. The cases $\Sigma=0$, $\Delta_\theta=0$, and
$\sin\theta=0$ are analyzed in the same way as the cubic NLED
Kerr-Newman-NUT-$\Lambda$ family discussed in Sec.~\ref{VA}. The case
$\Delta_r=0$, however, is modified by the nonlinear electromagnetic
parameter $\xi$, which changes the horizon structure. In this case,
$\Delta_r$ is a fourth-order polynomial in $r$, given by
\begin{equation}
\Delta_{r} =r^2-2mr+(1-n^2\Lambda)(a^2-n^2)-\frac{\Lambda r^2}{3}(r^2+a^2+6n^2)+(Q_e^2+Q_m^2)(1-\xi r^3),  \nonumber
\end{equation}
In this case, unlike the previous solution, $\Delta_{r}$ contains the term $- (Q_e^2+Q_m^2)\xi r^3$, introduced by the NLED parameter $\xi$, which renders the root analysis more intricate. Therefore, even in the case of $\Lambda=0$, there may exist three real roots, what gives three possible horizons, the inner, outer and cosmological. In the next subsection we analyze in detail the horizons that arise when $\Delta_{r}=0$
\subsection{Horizons of the Quartic NLED-Kerr-Newman-NUT generalization.}

We consider the fourth-order polynomial Eq. $\Delta_{r} (r) =0$ of the metric (\ref{m.POT4}),

\begin{equation}
\Delta_{r}(r) = -\frac{\Lambda}{3}r^4 - \xi Q^2r^3 + \left[1 - \frac{\Lambda}{3}(a^2 + 6n^2)\right]r^2 - 2mr + (1 - \Lambda n^2)(a^2 - n^2) + Q^2 = 0,
\end{equation}
whose positive real roots determine the horizons. The relations between the coeffcients and the roots of the polynomial are
\begin{equation}
    \begin{split}
        -\frac{3\xi Q^2}{\Lambda}&=r_1+r_2+r_3+r_4,\\
        a^2+6n^2-\frac{3}{\Lambda}&=r_1(r_2+r_3+r_4)+r_2(r_3+r_4)+r_3 r_4,\\
        -\frac{6m}{\Lambda}&=r_1 r_2 r_3+r_1 r_2 r_4+r_1 r_3 r_4+r_2 r_3 r_4,\\
        -3\left[\frac{(1 - \Lambda n^2)(a^2 - n^2) + Q^2}{\Lambda}\right]&=r_1 r_2 r_3 r_4.
    \end{split}
\end{equation}

The exact roots are given by Eqs. (\ref{rc12}) and (\ref{rc34}).
We analyze different cases:


\subsection{\texorpdfstring{Horizons of the Quartic NLED-Kerr-Newman-NUT spacetime for $\Lambda=0$}{}}

In the case $\Lambda = 0$ the equation determining the horizons is
\begin{equation}
\Delta_r = -\xi Q^2 r^3 + r^2 - 2mr + a^2 - n^2 + Q^2 = 0,
\end{equation}
where $Q^2 = Q_e^2 + Q_m^2$.
When $\xi = 0$, the inner and outer horizons reduce to those of the Kerr–Newman–NUT black hole:
\begin{equation}
r_\pm^{\text{KN-NUT}} = m \pm \sqrt{m^2 - (Q^2 + a^2 - n^2)}, \quad \text{with} \quad Q^2 + a^2 - n^2 \leq m^2.
\end{equation}

For $\xi \neq 0$, the roots are given by
\begin{widetext}
\begin{equation}
\frac{r_k}{m}= \frac{1}{3x} \left\{1 + 2\sqrt{1 - 6x} \cos\left[ \frac{1}{3} \arccos\left( \frac{2 - 18x+27x^2y}{2(1 - 6x)^{3/2}} \right) + \frac{2k\pi}{3} \right] \right\}, \quad x=m\xi Q^2, \quad y=\frac{Q^2 + a^2 - n^2}{m^2},
\end{equation}
\end{widetext}
with $k=0$ for the pseudocosmological horizon, $r_c$; $k=1$ for the inner horizon, $r_-$; and $k=2$ for the outer horizon, $r_+$.

To ensure real roots, the argument of the arccosine must be in the range $[-1,1]$, leading to the constraints:
\begin{equation}
mQ^2\xi \in \left[0, \frac{1}{6} \right], \qquad \frac{Q^2 + a^2 - n^2}{m^2} \leq \frac{4}{3},\qquad r_- \in [0, 2m], \quad r_+ \in [m, 3m], \quad r_c \in [2m, \infty).
\end{equation}

The extremal black hole occurs if $m \xi Q^2 = 1/6$ and $(a^2 - n^2 + Q^2) = 4m^2/3$, leading to a triple root at $r_+ = 2m$. This case is particularly interesting, as it mimics a Schwarzschild black hole but is sourced by a non-linear electromagnetic field. However, the NUT parameter reveals the true topologically non-trivial nature of the solution, since there is a conical singularity.

Moreover, the following inequality holds:
\begin{equation}
    r_+^{\text{KN}} \leq r_+^{\text{KN-NUT}} \leq r_+^{\text{NLE-KN-NUT}}.
\end{equation}

\begin{figure}[htbp] \centering
\subfloat[$a=0,~n=0$ and $\xi=0.11/m^3$.]{\label{f:1}\includegraphics[width=0.5\textwidth]{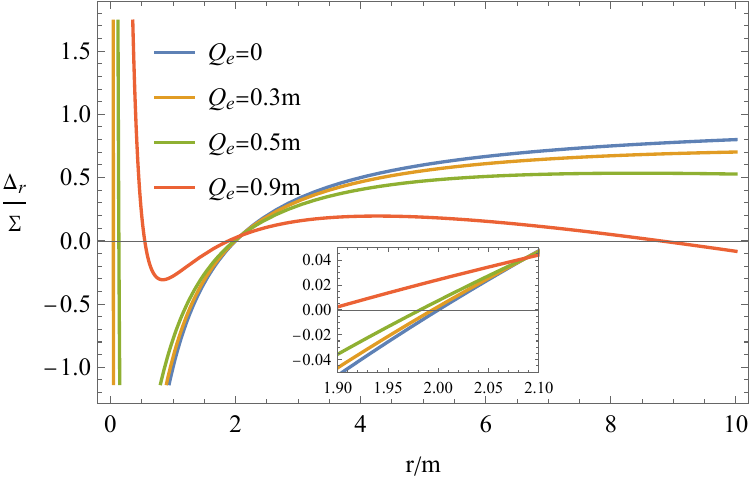}} \subfloat[$a=0,~n=0$ and $Q_e=0.5m$.]{\label{f:5}\includegraphics[width=0.5\textwidth]{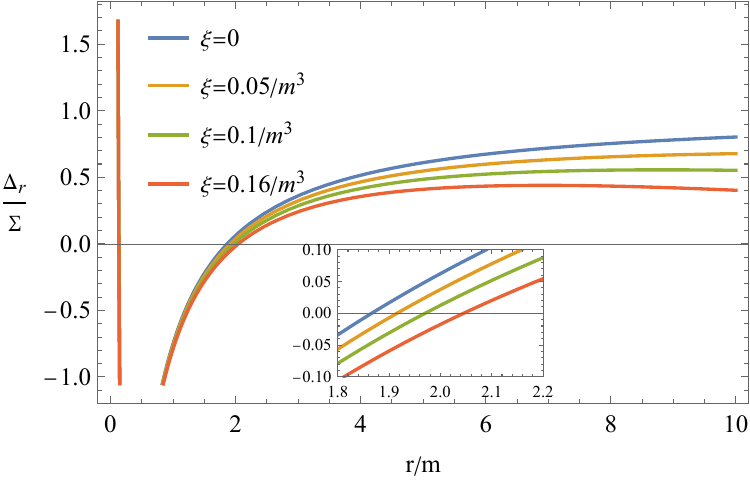}} \caption{ The Quartic NLED-RN metric function $\Delta_{r} / \Sigma$ is shown, with $a=0$ and $n=0$. In (a) is fixed $\xi=0.11/m^3$ while the charge $Q_e$ varies. In (b), the charge is kept constant at $Q_e=0.5m$ and the parameter $\xi$ increases. } \label{figNLED-RN} \end{figure}

In Figs. \ref{f:1} and \ref{f:5} the metric function $\Delta_{r}/ \Sigma $ is shown, first by fixing $\xi = 0.11/m^3$ and increasing $Q_e$ from $0$ to $0.9m$, the behavior of the radius of the event horizon is shown; it decreases monotonically. In contrast, fixing $Q_e = 0.5m$ and increasing $\xi$ from $0$ to $0.16/m^3$, the radius of the event horizon increases.

\subsubsection{\texorpdfstring{NLED-RN-NUT: $a = 0$, $n = 0.3m$}{}}

\begin{figure}[htbp] \centering \subfloat[$a=0,~n=0.3m$ and $\xi=0.11/m^3$.]{\label{ff:2}\includegraphics[width=0.5\textwidth]{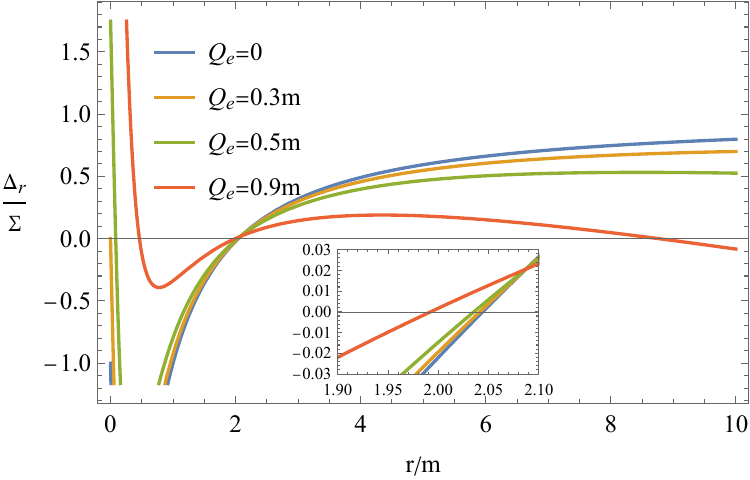}} \subfloat[$a=0,~n=0.3m$ and $Q_e=0.5m$.]{\label{ff:6}\includegraphics[width=0.5\textwidth]{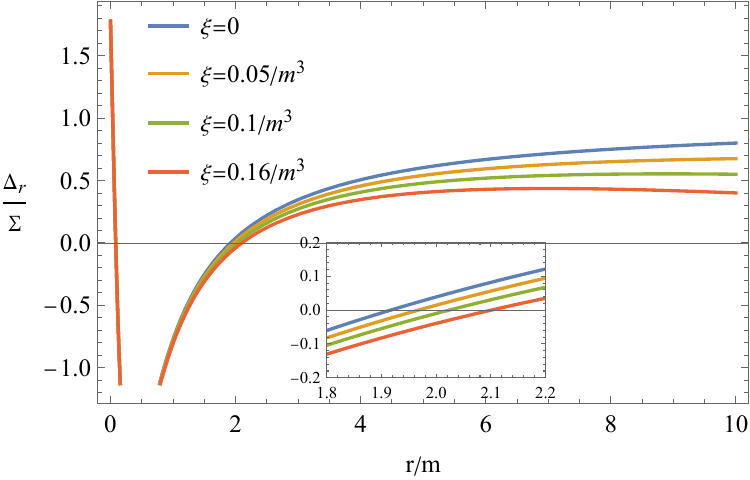}}\\ \caption{The Quartic NLED-RN-NUT metric function $\Delta_{r}/ \Sigma$ is shown, with $a=0$ and $n \neq 0$.
In (a) is fixed $\xi=0.11/m^3$ while the charge $Q_e$ varies. In (b) the charge is kept constant at $Q_e=0.5m$ and the parameter $\xi$ increases. In both figures $a=0$ and $n=0.3m$.} \label{figNLED-RN-NUT} \end{figure}

The quartic nonlinear electromagnetic generalization of the NLED-RN-NUT solution is a static one with a NUT parameter $n \neq 0$.
In Figs. \ref{figNLED-RN-NUT} the metric function $\Delta_{r}/ \Sigma $ of the NLED-RN-NUT solution is shown, first by fixing $\xi = 0.11/m^3$ and increasing $Q_e$ from $0$ to $0.9m$, with $a=0$ and $n=0.3m$, it is observed that the event horizon decreases. However, due to the presence of the NUT parameter $n$, the horizon is consistently larger than in the case $n = 0$ for all values of $Q_e$. Similarly, for fixed $Q_e = 0.5m$ and varying $\xi$ from $0$ to $0.16/m^3$, the horizon grows with $\xi$, and its magnitude is again larger than in the $n = 0$ case.

A special case arises when $(Q^2 + a^2 - n^2) = 0$ and $0 < Q \leq \frac{2m}{\sqrt{3}}$. The horizons are then given by
\begin{equation}
    r_0 = 0, \quad {\frac{r_\pm}{m}} = \frac{1 \pm \sqrt{1 - 8mQ^2\xi}}{2mQ^2\xi},
\end{equation}
subject to $0 < mQ^2\xi \leq \frac{1}{8}$, with
\begin{equation}
    r_- \in [0, 4m], \quad r_+ \in [4m, \infty).
\end{equation}

\begin{figure}[!ht]
\centering  \subfloat[$a=0.5m,~n=0.3m$ and $\xi=0.11/m^3$.]{\label{ff:4}\includegraphics[width=0.5\textwidth]{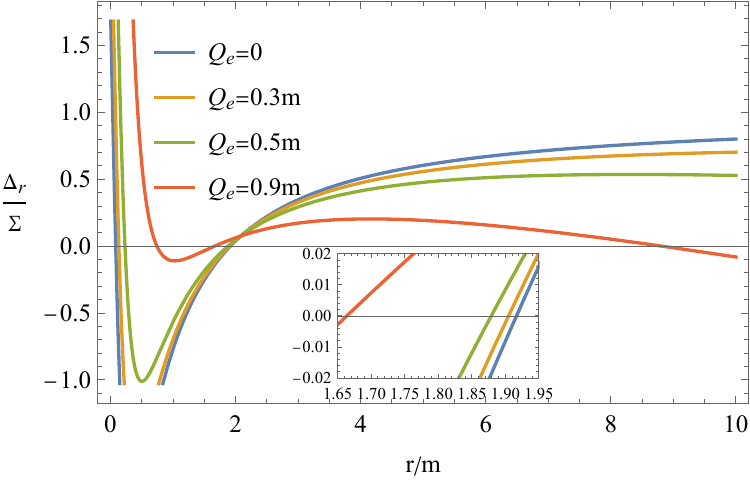}} \subfloat[$a=0.5m,~n=0.3m$ and $Q_e=0.5m$.]{\label{ff:8}\includegraphics[width=0.5\textwidth]{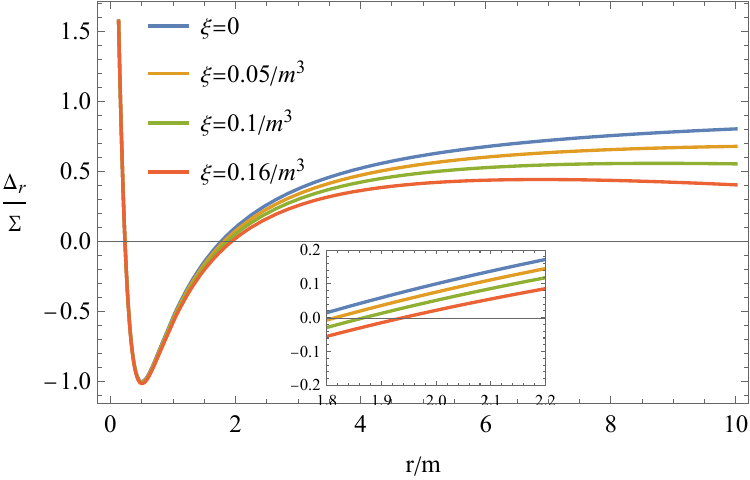}}\\ \caption{ The Quartic NLED-KN-NUT metric function $\Delta_{r}/ \Sigma$ is shown, with $a=0.5m$ and $n=0.3m$. In (a)  is fixed $\xi=0.11/m^3$ while the charge $Q_e$ varies. In (b), the charge is kept constant at $Q_e=0.5m$ and the parameter $\xi$ increases.} \label{figNLED-KN-NUT} \end{figure}

\begin{figure}[!ht] \centering  \subfloat[$a=0.5m,~n=0$ and $\xi=0.11/m^3$.] {\label{ff:3}\includegraphics[width=0.5\textwidth]{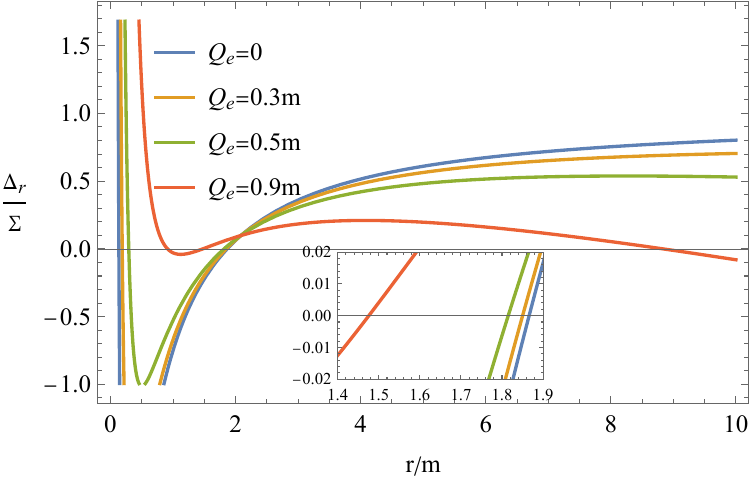}} \subfloat[$a=0.5m,~n=0$ and $Q_e=0.5m$.]{\label{ff:7}\includegraphics[width=0.5\textwidth]{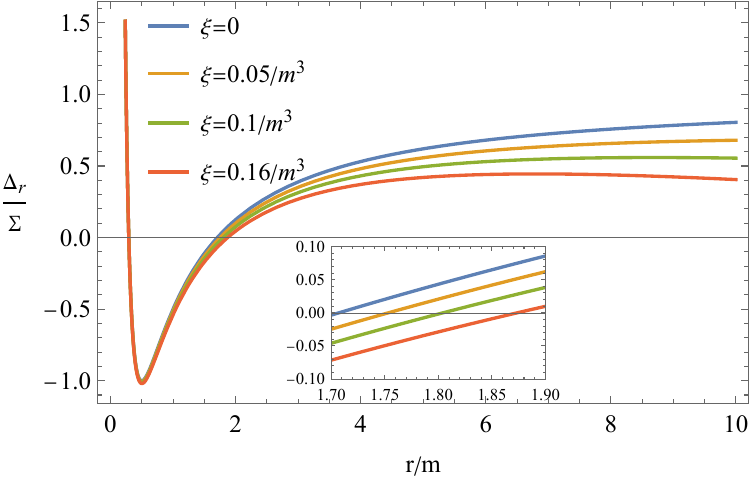}}\\ \caption{The Quartic NLED-KN metric function $\Delta_{r}/ \Sigma$ is shown, with $a=0.5m$ and $n=0$. In (a)  is fixed $\xi=0.11/m^3$ while the charge $Q_e$ varies. In (b), the charge is kept constant at $Q_e=0.5m$ and the parameter $\xi$ increases.} \label{figNLED-KN} \end{figure}

Several of these cases deserve a detailed analysis regarding their geodesics structure that promises interesting features.
The existence of horizons is determined by the values of the BH parameters.
In Figs.~ \ref{figNLED-KN-NUT} and  \ref{figNLED-KN} we display the metric function $\Delta_{r}/ \Sigma$ to show the effect on the event horizon of the variation of the parameters $Q_e$ and $\xi$ fixing $m$ as a scale.

The metric function $\Delta_{r}/ \Sigma$ for the Quartic NLED-Kerr-Newman-NUT solution, with $a = 0.5m$, $n = 0.3m$, is shown in
Figs. \ref{figNLED-KN-NUT} for $\xi = 0.11/m^3$ and $Q_e \in [0, 0.9m]$, the event horizon decreases with increasing $Q_e$. Besides, at $Q_e = 0$, we observe the ordering:
\begin{equation}
r_+(a = 0, n = 0.3m) > r_+(a = 0, n = 0) > r_+(a = 0.5m, n = 0.3m) > r_+(a = 0.5m, n = 0),
\end{equation}
indicating the combined effect of $a$ and $n$ on reducing or enlarging the horizon, respectively. This pattern is preserved when $Q_e = 0.5m$ is fixed and  $\xi$ is increased: the horizon grows with $\xi$, but remains the same hierarchy between configurations.

\subsubsection{\texorpdfstring{Case Quartic NLED-Kerr-Newman: $a = 0.5m$, $n = 0$}{}}

The metric function $\Delta_{r}/ \Sigma$ for the Quartic NLED-Kerr-Newman solution,
with $a = 0.5m$, $n = 0$, is displayed in Fig.\ref{figNLED-KN},
with $\xi = 0.11/m^3$ and  varying $Q_e$ from $0$ to $0.9m$; the radius of the event horizon decreases more significantly than in the non-rotating case, $a = 0$; this indicates that the spin parameter $a$ tends to shrink the event horizon. Similarly, for $Q_e = 0.5m$ and $\xi \in [0, 0.16]$, the horizon grows with $\xi$, and is larger than in the non-rotating case.

The electric charge $Q_e$ tends to reduce the event horizon; while the NLE parameter $\xi$ tends to increase it, because there is a charge screening effect. The NUT parameter $n$ consistently expands the horizon, while  increasing the rotation parameter $a$ consistently shrinks the horizon.  The final radius of the horizon is the result of the balance between these parameters.
\subsection{Quadruple degenerate horizon}

A mathematically admissible  scenario is a root of multiplicity four, which we denote by $r_+$. The quadruple degenerate root is defined by the conditions

\begin{equation}
\Delta_{r}(r_+) = 0, \quad \frac{d\Delta_{r}}{dr}(r_+) = 0, \quad \frac{d^2\Delta_{r}}{dr^2}(r_+) = 0, \quad \frac{d^3\Delta_{r}}{dr^3}(r_+) = 0.
\end{equation}
That occurs if the parameters satisfy the following relations:

\begin{equation}
\begin{split}
\frac{a^2}{m^2} &= \frac{2}{7}\left[3\frac{r_+^2}{m^2}\left(4 + \frac{r_+}{m}\right) - 2\sqrt{2\frac{r_+^3}{m^3}}\sqrt{\frac{r_+^3}{m^3}\left(21\frac{Q^2}{m^2} + 30\frac{r_+}{m} + 36\frac{r_+^2}{m^2} + 8\frac{r_+^3}{m^3}\right)}\right], \\
\frac{n^2}{m^2} &= \frac{1}{21}\left[\left(9 - 10\frac{r_+}{m}\right)\frac{r_+^2}{m^2} + 2\sqrt{2\frac{r_+^3}{m^3}}\sqrt{\frac{r_+^3}{m^3}\left(21\frac{Q^2}{m^2} + 30\frac{r_+}{m} + 36\frac{r_+^2}{m^2} + 8\frac{r_+^3}{m^3}\right)}\right], \\
\Lambda m^2 &= -\frac{3}{2}\frac{m^3}{r_+^3}, \quad m^3 \xi = \frac{2m^4}{Q^2 r_+^2}, \quad(1 - \Lambda n^2)(a^2 - n^2) + Q^2= \frac{r_+}{2}.
 \end{split}
\end{equation}

The quadruple root only exists in the interval $r_+ \in [0, 3m]$.
The previous constraints  in the BH parameters restrict their ranges
as follows,

\begin{equation}
a, n \in \left[0, \sqrt{\frac{8}{7}}\,m\right], \quad Q^2 \in (0, 1.5m], \quad \Lambda \in \left(-\infty, -\frac{1}{18m^2}\right], \quad \xi \in \left[\frac{4}{27m^3},\infty\right).
\end{equation}
The interest in this spacetime with only one horizon, although being a BH characterized by several parameters, is that it resembles a Schwarzschild BH
and, regarding observations, this case cannot be distinguished from the simplest BH defined by its mass.

\section{The nonlinear electromagnetic energy momentum tensor}\label{EnergyT}

To assess whether the NLE energy-momentum tensor $T_{\mu \nu}$  is physically reasonable, by this meaning that the local energy density measured by any observer is non-negative  as well as  the local energy flow vector to be non-spacelike, we check the energy conditions   associated to the NLE generalization of the KN-NUT-$\Lambda$.
To this end, we project  $T_{\mu \nu}$ onto the orthonormal tetrad, $\{\bf {E^a},a=1,\ldots,4\}$ associated to the metric (\ref{metric1}),  which is given by:

\begin{equation}
{\bf E^1} = \sqrt{\frac{\Sigma}{\Delta_\theta}} d\theta, \quad
{\bf E^2}  =  \frac{\sin\theta}{\Xi}\sqrt{\frac{\Delta_\theta}{\Sigma}}\left[adt-\left(\Sigma+a\chi\right)d\phi\right],\quad  {\bf E^3}=\sqrt{\frac{\Sigma}{\Delta_r}} dr,\quad
{\bf E^4} =\frac{1}{\Xi}\sqrt{\frac{\Delta_r}{\Sigma}}\left(dt-\chi d\phi\right).
\end{equation}

\begin{equation}
{ dt}=-\frac{\Xi \chi }{\sin\theta\sqrt{\Sigma \Delta_\theta}}{\bf E^2}+\frac{\Xi(\Sigma+a\chi)}{\sqrt{\Sigma\Delta_r}}{\bf E^4},\quad {dr}=\sqrt{\frac{\Delta_r}{\Sigma}}{\bf E^3},\quad { d\theta}=\sqrt{\frac{\Delta_\theta}{\Sigma}}{\bf E^1},\quad { d\phi}=-\frac{\Xi }{\sin\theta\sqrt{\Sigma \Delta_\theta}}{\bf E^2}+\frac{\Xi a}{\sqrt{\Sigma\Delta_r}}{\bf E^4}.
\end{equation}

Then, projecting  $T_{\mu \nu}$ onto the orthonormal basis, its canonical form is obtained,
\begin{equation}
\label{Orto.Tmunu}
OT^{ab}=
\begin{pmatrix}
\frac{T_{\theta \theta}}{g_{\theta \theta}}  & 0 & 0 & 0\\
0& \frac{T_{\theta \theta}}{g_{\theta \theta}} & 0 & 0 \\
0 & 0& \frac{T_{r r}}{g_{r r}}  & 0 \\
0 & 0& 0&  -\frac{T_{r r}}{g_{r r}}
\end{pmatrix}
=\begin{pmatrix}
        p_1&0&0&0\\
        0&p_2&0&0\\
        0&0&p_3&0\\
        0&0&0&\mu
    \end{pmatrix}.
\end{equation}
Where $p_1,~p_2,~p_3$ and $\mu$ are the eigenvalues of $T^{a b}$ and  represent the principal {\it pressures} in the three spacelike directions $\mathbf{E^\alpha}(\alpha=1,2,3)$, while $\mu$ represents the energy density  measured by an observer whose world-line at point $p$ has a unit tangent vector $\mathbf{E^4}$. Note that the $OT^{ab}$ tensor has the canonical form type I (\cite{HawkingEllis1973}).
In terms of electromagnetic potentials and  metric function
$\Delta_{r}(r)$ the $OT^{ab}$ components are

\begin{equation}
\label{Trr}
\mu =- \frac{T_{rr}}{g_{rr}}= -p_{3}= \frac{ED+BH}{8\pi}-\frac{{T}^{\mu}_{\mu}}{4}.
\end{equation}

\begin{equation}\label{T.theta.theta}
    p_{1,2} =\frac{ T_{\theta\theta}}{g_{\theta\theta}}= \frac{ED+BH}{8\pi}+\frac{{T}^{\mu}_{\mu}}{4}.
\end{equation}
\subsection{Energy conditions}
In what follows we determine the conditions on the $OT^{ab}$ components in order to fulfill the weak, dominant and strong energy conditions  that are summarized in Table \ref{table.EC}.
\begin{table}[H]
\centering
\begin{tabular}{|c | c c |c|}
 \hline
 Energy Cond. & $OT_{ab}$ components & & Inequalities\\
 \hline
 Weak (WEC) & $T_{ab}u^a u^b\geq 0$ & &  $\mu\geq0$, $\mu+p_\alpha\geq 0$ \\
 \hline
 Dominant (DEC)& $T_{ab}u^a u^b\geq 0$, $T^{ab}u_b \leq 0$ & & $\mu\geq 0$, $\mu\geq |p_\alpha|$\\
 \hline
\multirow{2}{*}{Strong  (SEC)}&\multirow{2}{*}{$(T_{ab}-\frac{1}{2}Tg_{ab})u^a u^b\geq 0$} && $\mu+\sum_{\alpha=1}^{3} p_\alpha\geq 0$,\\
& &    &$\mu+p_\alpha\geq 0$\\
\hline
\end{tabular}
\caption{Energy conditions. The inequalities that the stress-energy tensor
$T_{ab}$ should satisfy to fulfill the energy conditions; $u^{a}$  is a timelike vector; $\mu = - OT_{rr}/g_{rr}= -p_3$ and $p_1=p_2= OT_{\theta \theta}/g_{\theta \theta}$.}
\label{table.EC}
\end{table}


\subsection{\texorpdfstring{Cubic NLED-KN-NUT-$\Lambda$ solution: Energy Conditions}{}}

\subsubsection{Weak Energy Condition (WEC)}
If the energy momentum tensor is of type I, the Weak Energy Condition (WEC)\cite{HawkingEllis1973} holds if $\mu\geq0$. Additionally, the energy density should not be exceeded by any {\it pressure}, such that $\mu+p_{\alpha} \geq 0,\quad \alpha= 1, 2, 3$.
For the case of the cubic vector potential NLE-KN-NUT-$\Lambda$ solution, the Weak Energy Condition (WEC) amounts to the energy density $\mu$ restricted  to the following

\begin{equation}\label{POT3.WEC.1}
    \begin{split}
        0 \leq \mu &= \frac{(Q_e^2 + Q_m^2)}{8\pi\Sigma^2}(1 + \beta n^2)(1 + \beta n^2 + 3\beta r^2)(1 + \beta n^2 - \beta r^2),
\end{split}
\end{equation}
with the asymptotic behavior,
\begin{equation}
\lim_{r \to \infty} \mu = -\frac{3(Q_e^2 + Q_m^2)}{8\pi}(1 + \beta n^2)\beta^2; \quad
\lim_{r \to 0} \mu = \frac{(Q_e^2 + Q_m^2)(1 + \beta n^2)^3}{8\pi(a\cos\theta + n)^4}.
\end{equation}

Furthermore, the sum of the energy density $\mu$ and the pressures $p_{1,2}$ is given by:
\begin{equation}\label{POT3.WEC.2}
\begin{split}
(\mu + p_{1,2}) &= \frac{(Q_e^2 + Q_m^2)(1 + \beta n^2)}{4\pi \Sigma^2} \left\{1 + \beta \left[r^2 + 2n^2 - (n + a\cos\theta)^2\right] \right.\\
  &\left. \quad + \beta^2\left[(3r^2 - n^2)(n + a\cos\theta)^2\right] + n^2(r^2 + n^2) \right\}.
  \end{split}
\end{equation}
With the asymptotic behaviors,
\begin{equation}
\lim_{r \to \infty} (\mu + p_{1,2}) = 0; \quad
\lim_{r \to 0} (\mu + p_{1,2}) = \frac{(Q_e^2 + Q_m^2)}{4\pi(a\cos\theta + n)^4}(1 + \beta n^2)^2 \left\{1 - \beta \left[n^2 - (n + a\cos\theta)^2\right] \right\}.
\end{equation}

It is evident that when $\beta = 0$, both inequalities (\ref{POT3.WEC.1}) and (\ref{POT3.WEC.2}) are always satisfied, as expected for the Maxwell case, the KN-NUT-$\Lambda$ solution
.

If $n = 0$ in inequality (\ref{POT3.WEC.1}) then, as $r \to \infty$, the energy density $\mu$ becomes negative for any value of $\beta \neq 0$ due to the dependence $\sim -\beta^2$.  As $r \to 0$, the energy density $\mu$ is always positive. Therefore, there exists a critical radius $r_0$ such that $ \mu(r) \geq 0 \text{ for } r \in [0, r_0], \quad \mu(r) < 0 \text{ for } r > r_0.$ While if $n \neq 0$, then as $r \to \infty$, the energy density $\mu$ is positive only if $\beta \leq -\frac{1}{n^2}$. As $r \to 0$, the energy density $\mu$ is positive only if $\beta \geq -\frac{1}{n^2}$. Therefore, the weak energy condition (\ref{POT3.WEC.1}) is satisfied throughout the spacetime only if $\beta = -\frac{1}{n^2}$ or $\beta = 0$.

Hence, under these specific restrictions on the NLE parameter $\beta$, the energy density and its combinations with pressure hold the weak energy condition globally.

\subsubsection{ Dominant Energy Condition (DEC)}
The Dominant Energy Condition (DEC)\cite{HawkingEllis1973} holds if
$\mu\geq 0$, and $p_\alpha\leq |\mu|$. This leads to the following inequalities for the Cubic NLED-KN-NUT-$\Lambda$,
\begin{equation}
\label{POT3.DEC.1}
\begin{split}
  & 0 \leq ( \mu-p_{1,2})=-\frac{{T}^{\mu}_{\mu}}{2}=\frac{(Q_e^2+Q_m^2)}{4\pi \Sigma}(1+\beta n^2)(1+\beta n^2-3r^2\beta)\beta.
\end{split}
\end{equation}   
with the asymptotic behavior,
\begin{equation}
\lim_{r \to 0} (\mu-p_{1,2}) =\frac{(Q_e^2+Q_m^2)}{4\pi(a\cos\theta+n)^2} (1+\beta n^2)^2\beta; \quad
\lim_{r \to \infty} (\mu-p_{1,2}) =- 3\frac{(Q_e^2+Q_m^2)}{4\pi} (1+\beta n^2)\beta^2 .
\end{equation}
For $n = 0$, the DEC is satisfied globally only when $\beta = 0$. For $n \neq 0$, the DEC is globally satisfied if $\beta = -\frac{1}{n^2}$, which also guarantees the cancellation of the $\beta$-dependent terms at both asymptotic limits.
\subsubsection{ Strong Energy Condition (SEC)}
The SEC guarantees that matter is attractive, causing geodesics to converge. In terms of the $OT_{ab}$ components, SEC amounts to $\mu+p_1+p_2+p_3 \geq 0$.
For the cubic vector potential NLE-KN solution this reduces to
\begin{equation}\label{SEC.POT3}
\begin{split}
    0&\leq\frac{(Q_e^2+Q_m^2)}{4\pi\Sigma^2}(1+\beta n^2)\left\{1+2\beta \left[n^2-(a\cos\theta+n)^2\right]+\beta^2 \left[3r^4+n^4+2(n+a\cos\theta)^2(3r^2-n^2)\right]\right\},
\end{split}
\end{equation}
with the asymptotic behavior,
\begin{eqnarray}
&&\lim_{r \to 0} (\mu+p_1+p_2+p_3) = \frac{(Q_e^2+Q_m^2)}{4\pi(a\cos\theta+n)^4}(1+\beta n^2)^2\left\{1-\beta\left[2(n+a\cos\theta)^2-n^2\right]\right\}; \\
&& \lim_{r \to \infty} (\mu+p_1+p_2+p_3 )= \frac{3(Q_e^2+Q_m^2)(1+n^2\beta)\beta^2}{4\pi}.
\end{eqnarray}
The SEC is therefore globally satisfied only for $\beta = 0$ and $\beta = -\frac{1}{n^2}$, consistent with the analysis of the WEC and DEC.

The WEC, DEC, and SEC are globally satisfied throughout the spacetime only for the critical values $\beta = 0$ (linear Maxwell case) and $\beta = -\frac{1}{n^2}$. These values represent the physically admissible nonlinear electromagnetic generalization for which the spacetime avoids  violation of the classical energy conditions.
\subsection{\texorpdfstring{Quartic NLED-KN-NUT-$\Lambda$ solution: energy conditions}{}}
\subsubsection{Weak Energy Condition}

In the case of  the  NLE-KN-NUT-$\Lambda$ solution,  with a quartic vector potential and characterized by the NLE parameter $\xi$, WEC amounts to the following conditions,

\begin{equation}\label{POT4.WEC.1}
    \begin{split}
       0 \leq & \mu=\frac{(Q_e^2+Q_m^2)}{8\pi\Sigma^2}\left(1+2\xi r^3\right), \quad \mu+p_3=0.
\end{split}
\end{equation}
With the asymptotic behavior,
\begin{equation}
\lim_{r \to \infty} \mu = 0; \quad \lim_{r \to 0} \mu = \frac{(Q_e^2+Q_m^2)}{8\pi(a\cos\theta+n)^4}.
\end{equation}

For the combination $\mu + p_{1,2}$, we obtain

\begin{equation}\label{POT4.WEC.2}
\begin{split}
      0\leq& (\mu+p_{1,2})=\frac{(Q_e^2+Q_m^2)}{4\pi \Sigma^2}\left\{1+\frac{\xi r}{2}\left[r^2-3(a\cos\theta+n)^2\right]\right\},
\end{split}
\end{equation}
with the asymptotic behavior,
\begin{equation}
\lim_{r \to \infty} (\mu + p_{1,2}) = 0; \quad \lim_{r \to 0} \mu+p_{1,2} =\frac{(Q_e^2+Q_m^2)}{4\pi(a\cos\theta+n)^4} .
\end{equation}
Condition (\ref{POT4.WEC.1}) holds if the NLE parameter is greater than or equal to zero, $\xi \geq 0$. Condition (\ref{POT4.WEC.2}) is not automatically positive for all $r,\theta$. If $x=n+a\cos\theta$, the factor $1+\frac{\xi r}{2}\left(r^2-3x^2\right)$ has a minimum at $r=|x|$, with a value of $1-\xi {|x|}^3$. Therefore, to guarantee the global WEC, we need at least one additional condition of the form $0\leq\xi\left|n+a\cos\theta\right|^3\leq1$, and for all angles, by the sufficient bound

\begin{equation}\label{xi.cond}
     0\leq\xi\left(\left|n\right|+\left|a\right|\right)^3\leq1.
\end{equation} If $\xi=0$, the KN-NUT-$\Lambda$ solution is recovered, and the energy conditions are satisfied.
\subsubsection{ Dominant Energy Condition (DEC)}
The Dominant Energy Condition (DEC)\cite{HawkingEllis1973} holds if
$\mu\geq 0$, and $p_\alpha\leq |\mu|$. This leads to the following inequalities for the NLE-KN-NUT-$\Lambda$ solution

\begin{equation}\label{POT4.DEC.1}
    \begin{split}
       -(\mu+p_{1,2})\leq & 0\leq \mu-p_{1,2}=-\frac{\tensor{T}{^\mu_\mu}}{2}\\
   -\frac{(Q_e^2+Q_m^2)}{4\pi \Sigma^2}\left\{1+\frac{\xi r}{2}\left[r^2-3(a\cos\theta+n)^2\right]\right\} \leq & 0\leq\frac{3(Q_e^2+Q_m^2)}{8\pi \Sigma}\xi r,
\end{split}
\end{equation}
with the asymptotic behavior,
\begin{equation}
\lim_{r \to 0} -(\mu+p_{1,2}) =-\frac{(Q_e^2+Q_m^2)}{4\pi(a\cos\theta+n)^4}\leq 0\leq \lim_{r \to 0} \mu-p_{1,2} =0; \quad
\lim_{r \to \infty} -(\mu+p_{1,2}) =0\leq 0 \leq \lim_{r \to \infty} \mu-p_{1,2} =0.
\end{equation}
     
The case for $p_3$ reduces to $0\leq 2\mu$.

\subsubsection{ Strong Energy Condition (SEC)}
SEC guarantees that matter is attractive, causing geodesics to converge. In terms of the $OT_{ab}$ components SEC amounts to $(\mu+p_1+p_2+p_3) \geq 0$.
For the quartic vector potential NLE-KN-NUT-$\Lambda$ solution this reduces to

\begin{equation}\label{SEC.POT4}
\begin{split}
    0 & \leq (\mu+p_1+p_2+p_3)=\frac{(Q_e^2+Q_m^2)}{4\pi \Sigma^2}\left\{1-\xi r\left[r^2+3(a\cos\theta+n)^2\right]\right\},
\end{split}
\end{equation}
with the asymptotic behavior,
\begin{equation}
\lim_{r \to 0} (\mu+p_1+p_2+p_3) = \frac{(Q_e^2+Q_m^2)}{4\pi(a\cos\theta+n)^2}; \quad \lim_{r \to \infty} (\mu+p_1+p_2+p_3) = 0 .
\end{equation}
Although the asymptotic limit in Eq. (\ref{SEC.POT4}) vanishes as $r\to\infty$, it approaches zero from negative values for $\xi>0$. Thus, even when $\xi\geq 0$ and the bound $0\leq\xi\left(\left|n\right|+\left|a\right|\right)^3\leq1$ guarantee $\mu\geq0$ and the validity of the WEC and DEC inequalities in the corresponding angular domain, the SEC is not automatically satisfied for all $(r,\theta)$. In particular, the SEC requires
\begin{equation}
1-\xi r\left(r^2+3x^2\right)\geq0,
\quad x=n+a\cos\theta,
\end{equation}

Therefore, the SEC holds only inside the radial interval $r\in \left(0,\left(\frac{1+\sqrt{1+4x^6\xi^2}}{2\xi}\right)^{\frac{1}{3}}-\frac{x^2}{\left(\frac{1+\sqrt{1+4x^6\xi^2}}{2\xi}\right)^{\frac{1}{3}}}\right)$. Outside this interval, the combination $\mu+p_1+p_2+p_3$ becomes negative, although it tends to zero at infinity. Hence, the quartic model satisfies the energy conditions only in the corresponding admissible region of parameter space and spacetime, rather than globally for arbitrary $\xi>0$.


\section{Conclusions}\label{Conclusions}
We have constructed two exact aligned nonlinear-electrodynamic generalizations of the Kerr-Newman-NUT-$\Lambda$ spacetime, referred to here as the cubic and quartic families. These exact solutions of the Einstein-NLED equations are characterized by the physical parameters of the Kerr-Newman-NUT-$\Lambda$ geometry, namely mass, angular momentum, NUT parameter, cosmological constant, electric and magnetic charges and  one nonlinear electromagnetic parameter, $\beta$ or $\xi$, respectively. The construction follows the aligned-potential method introduced in Ref.~\cite{GarciaDiaz2022b} and described in Sec.~\ref{Method for generating NLED rotating solutions}. Within the polynomial aligned ansatz considered here, the key equation selects the admissible nonlinear branches, while the Einstein equations reduce to a single radial equation for the deformation $f(r)$ of the Kerr-like radial metric function $\Delta_r$.

The nonlinear parameters modify the horizon structure and may introduce an additional finite root that behaves as a pseudo-cosmological horizon. In the benchmark regimes displayed in the figures, increasing the electric charge tends to reduce the outer horizon radius, whereas increasing the nonlinear parameter tends to enlarge it. In the same parameter ranges, the NUT and rotation parameters act in competing ways: the NUT parameter tends to enlarge the horizon, while rotation tends to reduce it. The final horizon radius is therefore determined by the balance among the black-hole parameters. Several degenerate-horizon cases were also analyzed, yielding explicit constraints on the allowed parameter ranges.

The curvature analysis shows that the nonlinear electromagnetic terms do not generically remove the Kerr-like curvature singularity. The ring singularity is avoided only in the parameter regime $n^2>a^2$, for which $\Sigma\neq0$, nevertheless, the spacetime retains the conical singularity characteristic of the NUT spacetimes. Thus, the solutions obtained here are exact nonlinear-electrodynamic extensions of the Kerr-Newman-NUT-$\Lambda$ geometry, but they are not globally regular black holes. This conclusion is supported by the explicit Kretschmann scalars collected in Appendices \ref{KretS} and \ref{kret.quartic}.

We also determined the canonical form of the NLED energy-momentum tensor in an orthonormal tetrad and derived the  conditions for fulfillment of the weak, dominant, and strong energy-conditions. For the cubic family, the energy conditions are globally satisfied only in special parameter regimes, in particular in the Maxwell limit and in the critical NUT-supported case $\beta=-1/n^2$ with $n\neq0$. For the quartic family, $\xi\geq0$ is necessary for the positivity of the energy density. The remaining WEC and DEC inequalities impose the additional angular bound discussed in Eq.~\eqref{xi.cond}, while the SEC further restricts the radial domain and is therefore not satisfied throughout the whole spacetime. Hence, the energy conditions restrict the physically admissible domains of the parameters space.

The on-shell NLED Lagrangian was reconstructed as a function of the coordinates. In this form it consists of the Kerr-Newman-NUT-$\Lambda$ Maxwell contribution plus a nonlinear correction that vanishes when the NLED parameter is set to zero, thereby signaling the nonlinear electromagnetic effects. In selected static subsectors, the inversion to electromagnetic invariants can be performed explicitly, yielding Lagrangians in terms of one electromagnetic invariant, such as $\mathcal L(F)$. For the fully stationary rotating families, however, obtaining a closed global expression in terms of the invariants remains a nontrivial open problem. Moreover, the nonvanishing trace of the NLED energy-momentum tensor shows that the nonlinear electromagnetic sector breaks conformal invariance.

The present construction suggests that aligned-potential methods may be useful for exploring broader Plebański-type geometries, whose symmetries are inherited from the general type D Einstein-Maxwell solutions derived by Plebański in Ref.~\cite{Plebanski1975}. However, extending the construction beyond the polynomial aligned ansatz and beyond the dipolar electromagnetic structure assumed here remains an open problem. In particular, the potentials $A_\mu$ and $\tilde{\mathcal P}_\mu$ used in this work are not designed to generate more general electromagnetic multipolar structures.

Several directions deserve further study. The explicit electromagnetic fields obtained here provide a natural starting point for analyzing charged-particle trajectories around these NLED black holes. It would also be interesting to investigate in more detail the physical consequences of the degenerate-horizon constraints, the thermodynamic properties of the solutions, possible phase transitions signaled by divergences of the specific heat, and observational bounds on the nonlinear parameters from black-hole shadows and related strong-field observables. We leave these questions for future work.

{\bf Acknowledgements}
NB acknowledges partial support by SECIHTI-Mexico
project CBF2023-2024-811. The work of OG has been sponsored by Conahcyt-Mexico (SECIHTI-Mexico) through the Ph. D.  scholarship No. 815804. C.L. is
supported by the Deutsche Forschungsgemeinschaft
(DFG, German Research Foundation) under Germany’s
Excellence Strategy-EXC-2123 “Quantum Frontiers”
390837967, the CRC1464 “Relativistic and Quantum
based Geodesy” (TerraQ), and by the Research Training
Group GRK1620 “Models of Gravity”.
\appendix
\section{\texorpdfstring{The Kretschmann scalar for the Cubic NLED-KN-NUT-$\Lambda$}{The Kretschmann scalar for the Cubic NLED-KN-NUT-Lambda}}\label{KretS}

To investigate the regularity of the black hole solutions, we must determine the locus where its curvature invariants diverge. To this end, we calculate the Kretschmann scalar, defined by the full contraction of the Riemann tensor. Below, we present the exact Kretschmann scalar for the solution containing cubic NLED terms:
\begin{equation}
    \begin{split}
        \mathcal{K}_{\text{cubic}} &= \frac{8}{3}\left[\Lambda+\frac{(Q_e^2+Q_m^2)\beta(1+\beta n^2)(1+\beta n^2-3\beta r^2)}{\Sigma}\right]^2 \\
        &\quad +\frac{8(Q_e^2+Q_m^2)^2(1+\beta n^2)^2}{\Sigma^4}\left\{(1+\beta n^2)\left[1+\beta(r^2-n^2+x^2)\right]+3\beta^2x^2r^2\right\}^2 \\
        &\quad +\frac{48}{\Sigma^6}\Bigg\{n(r+x)(r^2-4rx+x^2)\left[1+\frac{\Lambda}{3}(a^2-4n^2)\right]-m(r-x)(r^2+4rx+x^2)\\
        &\quad +(1+\beta n^2)(Q_e^2+Q_m^2)\bigg[(r^2+2rx-x^2)\left[(1+\beta n^2)^2-\beta^2r^2x^2\right] \\
        &\quad +\beta(1+\beta n^2)\left(-\frac{r^4}{3}-2r^3 x+\frac{10}{3}r^2 x^2+2rx^3-\frac{x^4}{3}\right)\bigg]\Bigg\}\\
        &\quad \times \Bigg\{-n(r-x)(r^2+4rx+x^2)\left[1+\frac{\Lambda}{3}(a^2-4n^2)\right]-m(r+x)(r^2-4rx+x^2)\\
        &\quad +(1+\beta n^2)(Q_e^2+Q_m^2)\bigg[(r^2-2rx-x^2)\left[(1+\beta n^2)^2-\beta^2r^2x^2\right] \\
        &\quad +\beta(1+\beta n^2)\left(-\frac{r^4}{3}+2r^3 x+\frac{10}{3}r^2 x^2-2rx^3-\frac{x^4}{3}\right)\bigg]\Bigg\}.
    \end{split}
\end{equation}

The asymptotic behavior of the scalar at spatial infinity approaches a constant:
\begin{equation}
    \lim_{r\rightarrow\infty}\mathcal{K} = 24\left[\beta^2(1+\beta n^2)(Q_e^2+Q_m^2)-\frac{\Lambda}{3}\right]^2.
\end{equation}

\noindent Conversely, approaching the center of the coordinate system ($r \rightarrow 0$), the scalar behaves as:
\begin{equation}
\begin{split}
\lim_{r\rightarrow0}\mathcal{K} &= \frac{8\beta^2 (Q_e^2+Q_m^2)^2(1+\beta n^2 )^4}{x^4} +\frac{8}{3}\left[\Lambda+\frac{\beta (Q_e^2+Q_m^2)(1+\beta n^2 )^2}{x^2}\right]^2 \\
&\quad +\frac{48}{x^8}\left\{(Q_e^2+Q_m^2)(1+\beta n^2 )^2\left(1+\beta n^2 +\frac{\beta x^2}{3}\right)-x\left[\frac{\Lambda n}{3}\left(a^2-4n^2\right)+n+m\right]\right\} \\
&\quad \times \left\{(Q_e^2+Q_m^2)(1+\beta n^2 )^2\left(1+\beta n^2 +\frac{\beta x^2}{3}\right)-x\left[\frac{\Lambda n}{3}\left(a^2-4n^2\right)+n-m\right]\right\} \\
&\quad +\frac{8(Q_e^2+Q_m^2)^2(1+\beta n^2 )^4\left[(1+\beta n^2-\beta x^2)^2+\beta^2 x^4\right]}{x^8}.
\end{split}
\end{equation}

\section{\texorpdfstring{The Kretschmann scalar for the Quartic NLED-KN-NUT-$\Lambda$}{The Kretschmann scalar for the Quartic NLED-KN-NUT-Lambda}}\label{kret.quartic}

Recall from Eq.~(\ref{f.quartic}) that the radial deformation for the quartic family is
\begin{equation}
     f_{\text{quartic}}(r) = -\xi (Q_e^2+Q_m^2) r^3.
\end{equation}
The Kretschmann scalar in this case is given by:
\begin{equation}
    \begin{split}
        \mathcal{K}_{\text{quartic}} &= \frac{8}{3}\left[\Lambda+\frac{3\xi\left(Q_e^2+Q_m^2\right)r}{2\Sigma^2}\right]^2+\frac{8\left(Q_e^2+Q_m^2\right)^2}{\Sigma^4}\left[1+\frac{\xi r (r^2-3x^2)}{2}\right]^2 \\
        &\quad +\frac{48}{\Sigma^6}\Bigg\{(r+x)(r^2-4rx+x^2)\left[-\frac{\xi}{2}\left(Q_e^2+Q_m^2\right)rx+n+\frac{n\Lambda(a^2-4n^2)}{3}\right] \\
        &\quad -m(r-x)(r^2+4rx+x^2)+\left(Q_e^2+Q_m^2\right)(r^2+2xr-x^2)\Bigg\}\\
        &\quad \times \Bigg\{-(r-x)(r^2+4rx+x^2)\left[-\frac{\xi}{2}\left(Q_e^2+Q_m^2\right)rx+n+\frac{n\Lambda(a^2-4n^2)}{3}\right] \\
        &\quad -m(r+x)(r^2-4rx+x^2)+\left(Q_e^2+Q_m^2\right)(r^2-2xr-x^2)\Bigg\} .
    \end{split}
\end{equation}

Similarly, for the Quartic family, the asymptotic limits are found to be:
\begin{equation}
\lim_{r\rightarrow\infty}\mathcal{K} = \frac{8\Lambda^2}{3},
\end{equation}
and at the origin:
\begin{equation}
\begin{split}
\lim_{r\rightarrow0}\mathcal{K} &= \frac{48}{x^8}\left\{Q_e^2+Q_m^2-x\left[\frac{\Lambda n}{3}\left(a^2-4n^2\right)+n+m\right]\right\} \\ 
&\quad \times \left\{Q_e^2+Q_m^2-x\left[\frac{\Lambda n}{3}\left(a^2-4n^2\right)+n-m\right]\right\} \\
&\quad +\frac{8\Lambda^2}{3}+\frac{8(Q_e^2+Q_m^2)^2}{x^8}.
\end{split}
\end{equation}

Ultimately, both black hole solutions fail to be globally regular. The presence of $x^8$ in the denominators of the limit $r \rightarrow 0$ indicates that a curvature singularity persists. Since $x = n + a\cos\theta$, the condition $\Sigma = r^2 + x^2 = 0$ is satisfied at $r=0$ along the specific angular contour defined by $\cos\theta = -n/a$ (provided $|n| \leq |a|$). As one approaches this region, the Kretschmann scalar diverges ($\mathcal{K} \rightarrow \infty$). Therefore, we conclude that the introduction of these specific cubic and quartic NLED models modifies the internal geometry but is fundamentally insufficient to resolve the inherent ring singularity of the Kerr-Newman-NUT-$\Lambda$ spacetime.

\section{\texorpdfstring{On-shell Lagrangians and static invariant limits}{}}\label{app:Lagrangian}

The on-shell Lagrangian used throughout this work is obtained from the trace of the electromagnetic energy-momentum tensor. With the conventions adopted in Sec.~\ref{NLED-gravity Field equations}, it is given by

\begin{equation}
    \mathcal{L} =\frac{1}{4}\ {F}^{\mu\nu} P_{\mu \nu} +\pi {T}{^\mu_\mu} 
= \frac{ED-BH}{2} + \pi T^\mu_{\;\;\mu}.
\end{equation}

For a covariant nonlinear electrodynamics one would like to express the Lagrangian globally in terms of the electromagnetic invariants $F$ and $G$. For the fully rotating aligned families considered here, however, the map $(r,\theta)\longmapsto \bigl(F(r,\theta),G(r,\theta)\bigr)$
is not straightforwardly invertible in closed form. In particular, solving for $r=r(F,G)$ and $\theta=\theta(F,G)$ generally leads to algebraic equations of degree higher than four. Therefore, in the rotating case we present the on-shell Lagrangians in the intensity representation, namely in terms of $E$, $B$, $D$, and $H$. Whenever a simpler static subsector allows the inversion to be performed explicitly, we also display the corresponding invariant form.

\subsection{Nonlinear electromagnetic cubic Lagrangian}

For the cubic family, the Lagrangian can be written compactly as
\begin{equation}\label{LAG.POT3.general}
    \mathcal{L} = \frac{ED-BH}{2} + \frac{\beta}{2}\left\{ \frac{\left[(1+\beta n^2) \mathcal{X} - \mathcal{Y} + \frac{\beta (1+\beta n^2) Q^2}{2}\right]^2 - \beta (1+\beta n^2) Q^2 \mathcal{Y} +\frac{3}{4}\beta^2(1+\beta n^2)^2 Q^4}{(1+\beta n^2) \mathcal{X} - \mathcal{Y}} \right\},
\end{equation}
where
\begin{equation}
    \mathcal{X} = Q_m H - Q_e E, \quad \mathcal{Y} = Q_m B - Q_e D, \quad Q^2 = Q_e^2 + Q_m^2.
\end{equation}
This expression is an on-shell Lagrangian adapted to the aligned branch. Equivalently, one may attempt to reconstruct $\mathcal{L}(E,B)$ from the constitutive relations  $D=\frac{\partial \mathcal{L}}{\partial E}$ and $H=-\frac{\partial \mathcal{L}}{\partial B}$ by solving the corresponding differential equation for $\mathcal{L}(E,B)$.

In the general rotating case this reconstruction is not analytically tractable in a simple closed form. We consider the case where $a=0$ and $n=0$. On this static solution branch one also has, $E/D=H/B=\beta r^2+1$ while the constitutive equations reduce to
\begin{equation}
    D = E + \beta Q_e, \qquad H = B + \beta Q_m.
\end{equation}
Consequently, the Lagrangian of Eq.(\ref{LAG.POT3.general}) can be written in terms of the field intensities as
\begin{equation}
\begin{split}
    \mathcal{L} &= \frac{E^2-B^2}{2} +\beta Q_e E -\beta Q_m B+\frac{3}{2}\beta^2Q_m^2+2\beta^2 Q_e^2 .
\end{split}
\end{equation}

Expressing the Lagrangian purely in terms of the electromagnetic invariants $F$ and $G$, we obtain
\begin{equation}\label{Lpot3.a0n0}
\begin{split}
    \mathcal{L}&= -F +\beta Q_e \sqrt{\sqrt{F^2+G^2}-F} -\beta Q_m \sqrt{\sqrt{F^2+G^2}+F} +\frac{3}{2}\beta^2Q_m^2+2\beta^2 Q_e^2,
\end{split}
\end{equation}
where the field components are given by $E = \sqrt{\sqrt{F^2+G^2}-F}$ and $B = \sqrt{\sqrt{F^2+G^2}+F}$. Clearly, the nonlinear terms vanish when $\beta=0$. Furthermore, the expression simplifies significantly in the limits

\begin{align}
    a=0,~n=0,~Q_e&=0\leftrightarrow \mathcal{L}=-F\left(1+\beta Q_m\sqrt{\frac{2}{F}}\right)+\frac{3}{2}\beta^2Q_m^2,\\
    a=0,~n=0,~Q_m&=0\leftrightarrow \mathcal{L}=-F\left(1+\beta Q_e\sqrt{-\frac{2}{F}}\right)+2\beta^2 Q_e^2.
\end{align}

\subsection{Nonlinear electromagnetic quartic Lagrangian}

For the quartic family, the on-shell Lagrangian is
\begin{subequations}
\begin{align}
    \mathcal{L} &= \frac{ED-BH}{2} + \frac{3}{4} \xi^2 (Q_e^2+Q_m^2)^2 \left( \frac{\Psi_1}{\Psi_1^2 + \Psi_2^2} \right), \label{L_quartic_compact}
\end{align}
where the auxiliary functions $\Psi_1$ and $\Psi_2$ are defined in terms of the field intensities as
\begin{align}
    \Psi_1 &= n^3\xi(Q_e H+Q_m E) - 2Q_e(D-E) + 2Q_m(B-H) = -(Q_e^2+Q_m^2)\xi r, \label{Psi1_def} \\
    \Psi_2 &= n^3\xi(-Q_e D+Q_m B) + 2Q_e(B-H) + 2Q_m(D-E) = -(Q_e^2+Q_m^2)\xi (n+a\cos\theta). \label{Psi2_def}
\end{align}
\end{subequations}

An expression depending solely on the field intensities can be obtained when $a=0$ and $n=0$, yielding
\begin{equation}
    \mathcal{L} = \frac{ED-BH}{2} - 3\left(\frac{D^2+B^2}{2}\right)\left(\frac{E}{D}-1\right) = \frac{ED-BH}{2} - 3\left(\frac{D^2+B^2}{2}\right)\left(\frac{H}{B}-1\right).
\end{equation}

This expression displays explicitly the nonlinear deviation from the Maxwell Lagrangian within the static branch.  Furthermore, the expression simplifies significantly in the limit $a=0,~n=0,~Q_e=0$,

\begin{align}
  \mathcal{L}=-F-2^{\frac{1}{4}}\xi Q_m^{\frac{3}{2}}F^{\frac{1}{4}}.
\end{align}

\section{\texorpdfstring{Horizons of the NLED-KN-NUT-$\Lambda$ Spacetime}{}}\label{app:roots}

The horizons of the nonlinear electrodynamics (NLED) generalization of the Kerr–Newman–NUT–$\Lambda$ (KN-NUT-$\Lambda$) spacetime are determined by the condition $\Delta_r = 0$, which results in a quartic equation of the form
\begin{equation}
    a_4 r^4 + a_3 r^3 + a_2 r^2 + a_1 r + a_0 = 0,
\end{equation}
where $a_0, a_1, a_2, a_3, a_4 \in \mathds{R}$ and $a_4 \neq 0$. The general solution to a quartic equation yields four roots, given by

\begin{align}
    \label{rc12}
    r_{1_+,2_-} &= -\frac{a_3}{4a_4} + \frac{1}{2} \sqrt{ \frac{a_3^2}{4a_4^2} - \frac{2a_2}{3a_4} + \mathcal{B} }
    \pm \frac{1}{2} \sqrt{ \frac{a_3^2}{2a_4^2} - \frac{4a_2}{3a_4} - \mathcal{B}
    + \frac{ -\frac{a_3^3}{4a_4^3} + \frac{a_2 a_3}{a_4^2} - \frac{2a_1}{a_4} }{ \sqrt{ \frac{a_3^2}{4a_4^2} - \frac{2a_2}{3a_4} + \mathcal{B} } } },
    \\
    \label{rc34}
    r_{3_+,4_-} &= -\frac{a_3}{4a_4} - \frac{1}{2} \sqrt{ \frac{a_3^2}{4a_4^2} - \frac{2a_2}{3a_4} + \mathcal{B} }
    \pm \frac{1}{2} \sqrt{ \frac{a_3^2}{2a_4^2} - \frac{4a_2}{3a_4} - \mathcal{B}
    - \frac{ -\frac{a_3^3}{4a_4^3} + \frac{a_2 a_3}{a_4^2} - \frac{2a_1}{a_4} }{ \sqrt{ \frac{a_3^2}{4a_4^2} - \frac{2a_2}{3a_4} + \mathcal{B} } } }.
\end{align}

Here, the auxiliary function $\mathcal{B}$ is defined as
\begin{align}
    \mathcal{B} = \frac{2}{3} \sqrt{ \frac{a_2^2 - 3a_3 a_1 + 12a_4 a_0}{a_4^2} }
    \cos\left\{ \frac{1}{3} \arccos\left[
    \frac{ 2a_2^3 - 9a_2(a_1 a_3 + 8a_0 a_4) + 27(a_0 a_3^2 + a_1^2 a_4) }
    { 2 \left( a_4 \sqrt{ \frac{a_2^2 - 3a_3 a_1 + 12a_4 a_0}{a_4^2} } \right)^3 }
    \right] \right\}.
\end{align}

\bibliography{ref.bib}

\end{document}